\documentclass[10pt,journal,compsoc]{IEEEtran}

\ifCLASSOPTIONcompsoc
  \usepackage[nocompress]{cite}
\else
  \usepackage{cite}
\fi
\usepackage{makecell}
\usepackage{color}
\usepackage{multirow}
\usepackage{graphicx}
\usepackage{subfig}
\usepackage{algorithmic}
\usepackage{amsmath}
\usepackage{hyperref}
\usepackage[ruled,vlined]{algorithm2e}

\usepackage{booktabs}
\usepackage{subfig}
\newcommand{\jing}{\textcolor{black}}
\newcommand{\minor}{\textcolor{black}}
\newcommand{\namex}{Graft\xspace}
\newcommand{\mypara}[1]{\noindent\textbf{#1}}
\begin{document}
\title{\namex: Efficient Inference Serving for Hybrid Deep Learning with SLO Guarantees via DNN Re-alignment}

\author{Jing~Wu,
        Lin~Wang,~\IEEEmembership{Senior~Member,~IEEE},
        Qirui~Jin,
        Fangming~Liu\IEEEauthorrefmark{1},~\IEEEmembership{Senior~Member,~IEEE}
\IEEEcompsocitemizethanks{
\IEEEcompsocthanksitem This work was supported in part by National Key Research \& Development (R\&D) Plan under grant 2022YFB4501703, and the Major Key Project of PCL under Grant PCL2022A05. Lin Wang was supported in part by the Deutsche Forschungsgemeinschaft (DFG, German Research Foundation) – Project-ID 210487104 - SFB 1053. (Corresponding  author: Fangming Liu)
\IEEEcompsocthanksitem J. Wu and Q. Jin are with the National Engineering Research Center for Big Data Technology and System, the Services Computing Technology and System Lab, Cluster and Grid Computing Lab in the School of Computer Science and Technology, Huazhong University of Science and Technology, 1037 Luoyu Road, Wuhan 430074, China. E-mail: wujinghust@hust.edu.cn and qiruijin@umich.edu.
\IEEEcompsocthanksitem L. Wang is with Paderborn University and TU Darmstadt, Germany. E-mail: lin.wang@uni-paderborn.de.
\IEEEcompsocthanksitem F. Liu is with Peng Cheng Laboratory, and Huazhong University of Science and Technology, China. E-mail: fangminghk@gmail.com.
}}

%
%

\markboth{IEEE Transactions on Parallel and Distributed Systems,~Vol.~xx, No.~x, xxxx~xxxx}%
{Wu \MakeLowercase{\textit{et al.}}: HiTDL: High-Throughput Deep Learning Inference at the Hybrid Mobile Edge}
%

\IEEEtitleabstractindextext{%
\begin{abstract}
Deep neural networks (DNNs) have been widely adopted for various mobile inference tasks, yet their ever-increasing computational demands are hindering their deployment on resource-constrained mobile devices.
Hybrid deep learning partitions a DNN into two parts and deploys them across the mobile device and a server, aiming to reduce inference latency or prolong battery life of mobile devices.
However, such partitioning produces (non-uniform) DNN fragments which are hard to serve efficiently on the server.

This paper presents \namex---an efficient inference serving system for hybrid deep learning with latency service-level objective (SLO) guarantees.
Our main insight is to mitigate the non-uniformity by a core concept called DNN re-alignment, allowing multiple heterogeneous DNN fragments to be restructured to share layers.
To fully exploit the potential of DNN re-alignment, \namex employs fine-grained GPU resource sharing. 
Based on that, we propose efficient algorithms for merging, grouping, and re-aligning DNN fragments to maximize request batching opportunities, minimizing resource consumption while guaranteeing the inference latency SLO.
We implement a \namex prototype and perform extensive experiments with five types of widely used DNNs and real-world network traces. Our results show that \namex improves resource efficiency by up to 70\% compared with the state-of-the-art inference serving systems.
\end{abstract}

}

\maketitle

\IEEEdisplaynontitleabstractindextext

\IEEEpeerreviewmaketitle

\section{Introduction}
\label{sec:introduction}
                                                      
With the rapid development of deep learning (DL), we have seen a variety of DL-based mobile applications such as augmented reality~\cite{asplos18-potluck,mobicom21-smart,mobicom21-rfid} and intelligent personal assistant~\cite{socc20-inferline}. 
These applications typically employ pre-trained deep neural networks (DNNs) to perform inference tasks such as object detection and natural language understanding.
While achieving unprecedentedly high accuracy, modern DNNs are bloated in size and impose excessive computational demands~\cite{2019-csur-dl,2021-atc-infaas}.
Meanwhile, mobile inference requires low latency, typically specified with a service-level objective (SLO), to ensure good user experience~\cite{nsdi17-clipper,2019-sosp-nexus,mobicom20-spinn,usenix20-clockwork,tecs22-dyno}. 
Recent advances in mobile accelerators and DNN optimizations mitigate this mismatch issue, but bring concerns over battery life or incur extra costs from model retraining~\cite{imc21-mobile-measurement,mlsys22-mlperf-mobile,iccvw19-ai-benchmark,daes22-eanble-dl-on-mobile}.


\begin{figure}[!t]
	\centering
	\includegraphics[page=1,width=0.45\textwidth]{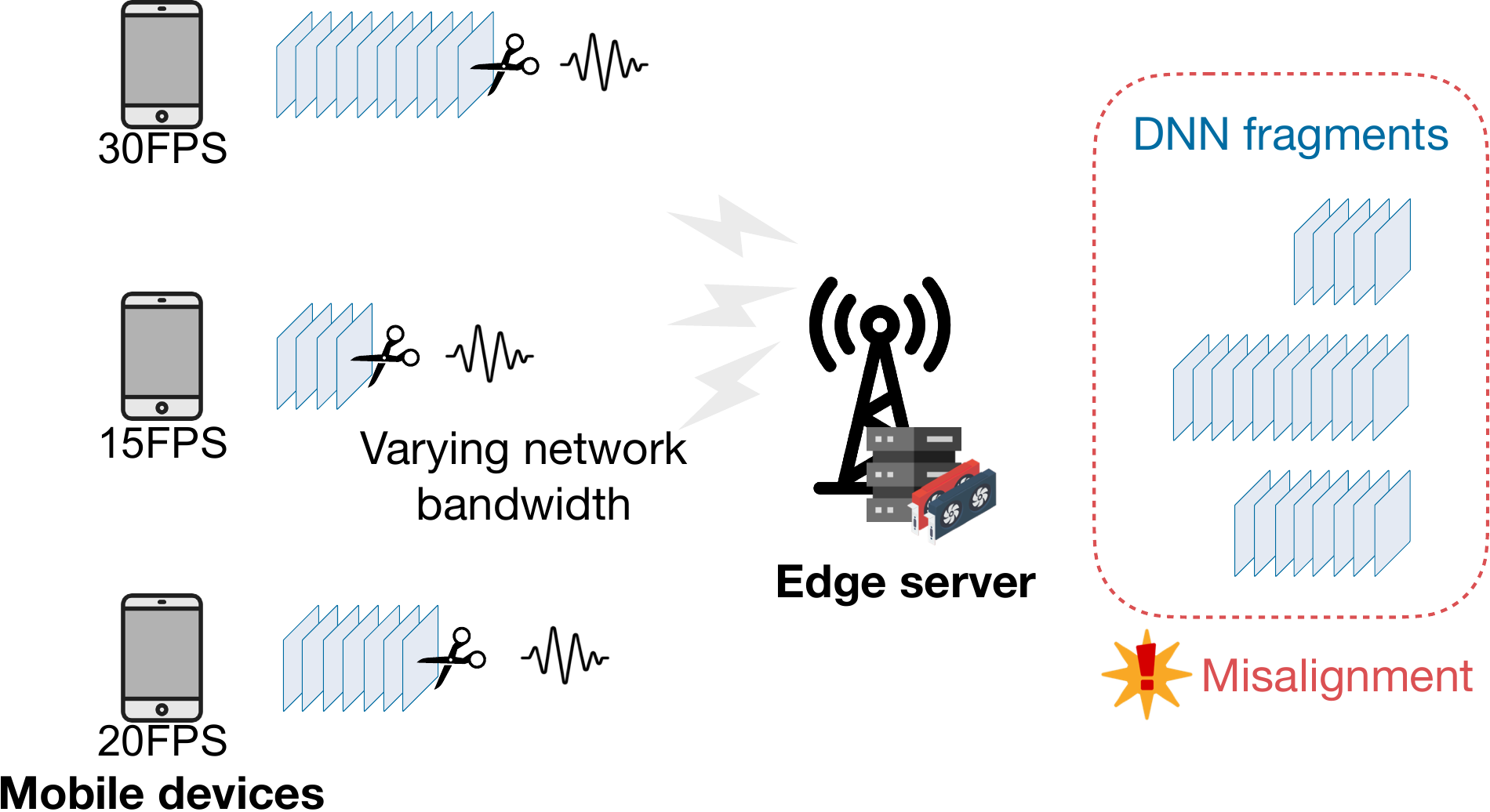}
	\caption{Hybrid deep learning across the mobile device and the edge server, resulting in misaligned DNN fragments on the edge server, which are hard to serve efficiently.\label{fig:hdl}}
	\vspace{-0.5cm}
\end{figure}

Hybrid DL tackles this challenge by extending the mobile device with servers typically in an edge cloud~\cite{cikm20-edgerec,arxiv21-mcsf, sigkdd21-autosplit,pytorch-glow,aws-neuron,aws-sagemaker,msr-fiddle,mobicom20-spinn,asplos17-neurosurgeon,mobicom20-clio,www21-auto-neurosurgeon}. 
Specifically, hybrid DL partitions a DNN into two parts (referred to as fragments) and executes them on the mobile device and the server, respectively, as depicted in Figure~\ref{fig:hdl}.
Unlike server-only solutions where the DNN is offloaded completely to a server, hybrid DL is still able to exploit the computing resources available on the mobile device for partial DNN execution and can accommodate network dynamics through adaptive DNN partitioning~\cite{asplos17-neurosurgeon}. 
Hybrid DL has been widely explored in both academia~\cite{arxiv18-jointdnn,mobicom20-spinn,asplos17-neurosurgeon,tpds22-hitdl} and industry~\cite{cikm20-edgerec,sigkdd21-autosplit} for minimizing the inference latency or maximizing the battery life of mobile device. 
It has been shown that hybrid DL can reduce the inference latency of popular DNNs by up to $1.69\times$~\cite{ubicomp20-qdmp}.
\jing{In addition, hybrid DLs offer enhanced privacy protection since they only exchange intermediate data, which blurs a significant level of personal information due to local inferences~\cite{asplos20-shredder,tecs22-dyno,corr19-split-learning}.}

Despite the vast successes, efficiently serving the generated DNN fragments on the server side remains a critical challenge, which has been overlooked so far. 
In particular, different mobile devices running the same DNN may produce varying partition decisions due to their different yet changing network conditions.
Consequently, the server will need to handle a large number and variety of \emph{misaligned} DNN fragments with relatively small inference request rates from individual mobile devices, as depicted in Figure~\ref{fig:hdl}.
Although these DNN fragments share layers partially, the lack of uniformity due to the misalignment largely limits the opportunity of applying request batching---an essential technique for improving resource efficiency in virtually all inference serving systems~\cite{usenix20-clockwork}.
\jing{Existing inference serving systems are either agnostic to such partial sharing and would simply treat these DNN fragments separately, or blindly merge architecture-identical fragments without further considering other individual requirements (e.g., time budget) for servers, leading to poor resource efficiency that is detrimental to edge clouds.}

\jing{This paper addresses this timely challenge by introducing the concept of \emph{DNN re-alignment} for hybrid DL.
The key idea is to re-partition, i.e., apply to another partitioning, non-uniform DNN fragments such that we can generate aligned DNN fragments that consist of the same DNN layers.
These aligned DNN fragments can then be served by shared DNN instances where request batching can be applied as in existing DL inference serving systems.
By deciding the re-partition point carefully, we can leverage request batching to its full potential, thus improving the server resource efficiency in hybrid DL.
The entire optimizations on non-uniform fragments for higher resource efficiency refer to re-alignment.}

We propose \namex---a first-of-its-kind inference serving system adopting DNN re-alignment for hybrid DL. \emph{The goal of \namex is to achieve resource efficiency while guaranteeing latency SLO}. 
To this end, \namex re-aligns the DNN fragments on the server to enable request batching across inference requests received from different mobile devices.
\namex features efficient strategies for merging and grouping DNN fragments by converting the grouping problem into a variant of the balanced graph partitioning problem.
\namex also incorporates a fine-grained resource allocation policy based on spatial GPU sharing enabled by CUDA MPS (Multi-Process Service) of NVIDIA GPUs.
Specifically, \namex formulates joint re-alignment and resource allocation as an optimization problem and produces near-optimal decisions on the re-partition point, batch size, GPU allocation per instance, and the number of instances for all fragments on the server.
Ultimately, \namex minimizes the GPU resource consumption on the server while ensuring that the end-to-end latency for each inference request is bounded by the SLO.

Overall, the paper makes the following contributions: after identifying the challenges of provisioning misaligned DNN fragments in hybrid DL (\S\ref{sec:background}), we
\begin{itemize}
	\item introduce a new concept called DNN re-alignment and present the design of \namex---a first-of-its-kind inference serving system tailored for hybrid DL with latency SLO guarantees, employing DNN re-alignment (\S\ref{sec:system}).
	\item present efficient algorithms for DNN re-alignment, featuring greedy DNN fragment merging, graph-partitioning based fragment grouping, and fragment re-partitioning and resource allocation, with near-optimal resource efficiency (\S\ref{sec:scheduling}).
	\item implement \namex and perform extensive experiments with five different DNNs to evaluate \namex based on a system prototype (\S\ref{sec:evaluation}). Experiment results show that \namex is able to achieve resource savings by up to 70\% when compared with the state-of-the-art inference serving systems, while guaranteeing latency SLOs.
\end{itemize}
\S\ref{sec:relatedwork} discusses related work and \S\ref{sec:conclusion} draws the conclusions.

\section{Background and Motivation}
\label{sec:background}

We present the background and discuss the challenges and insights to motivate our work in this section.

\subsection{Deep Learning Deployment}
Deep learning (DL) has become the de facto approach for inference tasks in domains like computer vision and natural language processing~\cite{socc20-inferline,asplos17-neurosurgeon,axiv17-mobilenet}.
To improve accuracy, sophisticated DNNs with a large number of parameters have been explored~\cite{hotos21-dnn}, requiring substantial computing resources to execute.
\jing{For example, the computing demand (GFLOPS) of image classification DNNs has increased by up to 595.6$\times$ since 2016~\cite{paper-with-code-gflops} and the memory requirement of DNNs for high-resolution image transformations has reached 6-8GB~\cite{iccvw19-ai-benchmark}.}
This poses critical challenges for DL deployment, especially on resource-constrained mobile devices~\cite{asplos17-neurosurgeon,mobisys22-tinynet,mobicom21-high-resolution-detection}.

One solution is to fit the DNN execution on mobile devices via DNN optimizations including pruning and quantization ~\cite{iclr20-ofa,eccv18-pruning,nips19-bit,agi17-specialization,mobisys22-codl,mobisys22-mgemm, mobicom21-aysmo,mobicom21-legodnn}. 
\jing{Without optimizations, a DNN for image segmentation can drain the 4000mAh battery of a mobile device by 95.9\% in just one hour when running on mobile CPUs~\cite{imc21-mobile-measurement}.} 
DNN optimizations reduce the computational demands but suffer from an accuracy gap and lack practical deployment support. 
For example, empirical studies show that only 10.3\% of the DNN-based apps from the Google Play store employ DNN optimizations~\cite{imc21-mobile-measurement}. 
Developers tend to build customized DNNs for their apps with lower resource consumption but at the cost of lower accuracy~\cite{www19-first-look-mobile-apps}. 
The associated costs for optimizing and customizing a DNN for diverse mobile devices are considerable~\cite{iclr20-ofa,sigmetrics21-nas}. 
Meanwhile, over 50\% of the existing mobile devices still use processors that are at least six years old~\cite{hpca19-inference-facebook}, which clearly falls short of meeting the resource requirements of modern DNNs in entirety. 

Another solution is to offload the DNN execution completely to a powerful server equipped with accelerators like GPUs, typically at the network edge over a wireless network~\cite{computer17-edge,eurosys19-grandslam,atc22-soter}. 
This approach avoids the needs for DNN optimizations thanks to the high capability of the server, but it brings two other issues (as illustrated in Figure~\ref{fig:bg:partition_point} (top)): 
(1) Leaving the resources on mobile devices unused leads to excessive resource consumption on the server. 
(2) The inference latency suffers from uncertainty due to the network variability, leading to potentially high SLO violations.

\subsection{Hybrid Deep Learning}

To overcome these limitations, \emph{hybrid DL} aims to leverage computing resources on both the mobile device and the edge server for DL inference, without sacrificing on accuracy.
Considering the layered structure of DNNs, the key idea is to partition the DNN into two fragments at an intermediate layer, with each fragment mapped to the mobile device and to the server, respectively \cite{asplos17-neurosurgeon,sec19-deepsave,infocom19-dads,mobicom20-spinn,hotedgevide19-split-brain}.
Hybrid DL has been widely explored by the academic (e.g., Neurosurgeon~\cite{asplos17-neurosurgeon} and SPINN~\cite{mobicom20-spinn}) and the industry (e.g., Taobao~\cite{cikm20-edgerec,arxiv21-mcsf} and Huawei Cloud~\cite{sigkdd21-autosplit,huawei-cloud-edge-cloud-synergy}), with different partitioning strategies proposed to minimize the inference latency or maximize the battery life of the mobile device.
Recently, it has been shown that hybrid DL can achieve a speedup of 1.69$\times$ and reduce mobile energy consumption by 22.5\% on average~\cite{ubicomp20-qdmp,infocom22-aodnn}.

\begin{figure}[!t]
	\centering
	\centering
	\includegraphics[width=0.43\textwidth]{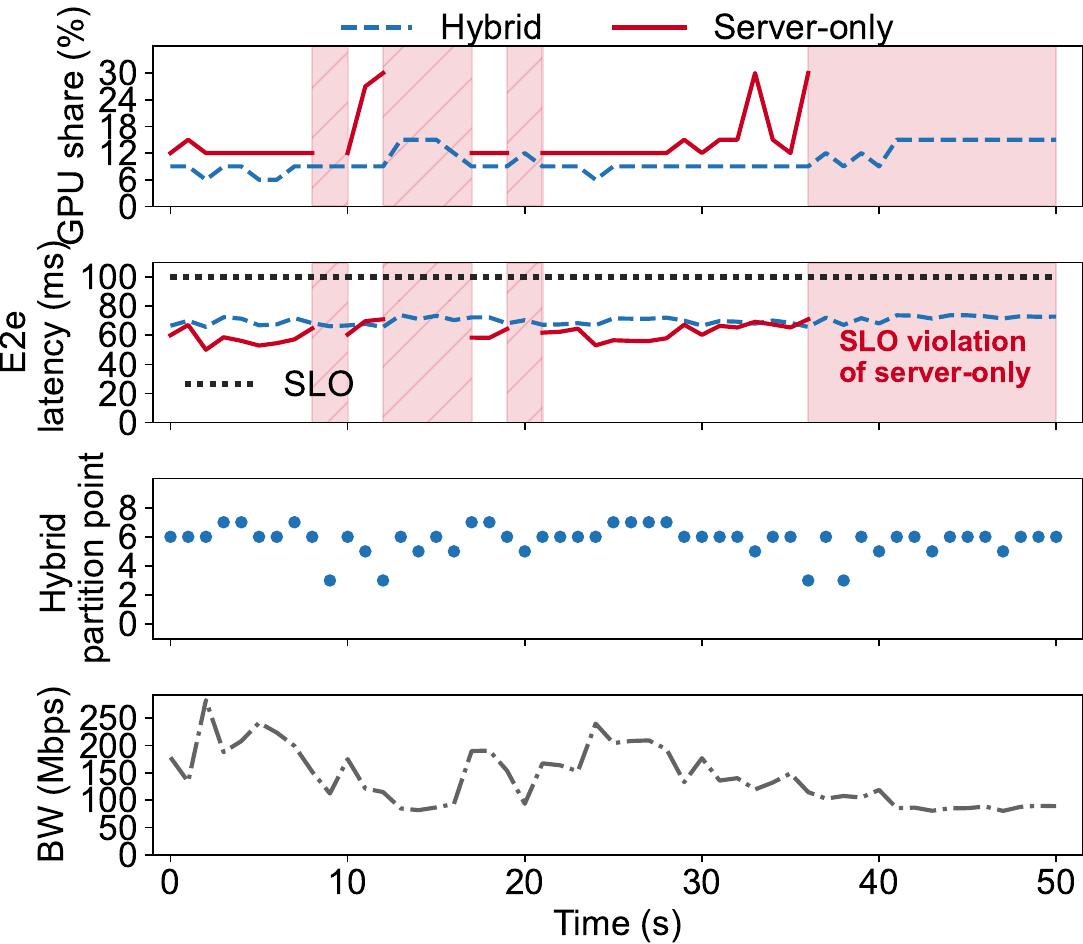}
	\caption{Changing resource consumption (top) and DNN partition points (middle) of Inception-v3 in hybrid DL under a snippet (50s) of a real-world 5G network trace~\cite{mmsys20-5g}.}
	\label{fig:bg:partition_point}
\end{figure}

Since the network bandwidth is likely to be dynamic, DNN partitioning thus needs to be adapted continuously. 
Figure~\ref{fig:bg:partition_point} (middle) shows the partition point variation under a real-world 5G network bandwidth trace (Figure~\ref{fig:bg:partition_point}(bottom))~\cite{mmsys20-5g} when applying hybrid DL. 
\minor{We can see from Figure~\ref{fig:bg:partition_point}(top) that hybrid DL accommodates the variability of network bandwidth; meanwhile,
by exploiting the resources on mobile devices, hybrid DL can reduce the server resource consumption by up to 3$\times$ and avoids SLO violations via dynamic DNN partitioning.}
So far, the focus of existing work has been on the partitioning strategy and little attention has been paid to the efficient serving of the DNN fragments on the edge server.

\subsection{DL Inference Serving for Hybrid DL}

The proliferation of DL-based applications renders the critical importance of DL inference serving.
There exist a large variety of DL inference serving systems, such as
TensorFlow Serving~\cite{tensorflow-serving}, Clipper~\cite{nsdi17-clipper}, Nexus~\cite{2019-sosp-nexus}, Clockwork~\cite{usenix20-clockwork}, ALERT~\cite{usenix20-alter}, and INFaaS~\cite{2021-atc-infaas}, etc., that concern scheduling DL inference requests to appropriate resources (e.g., GPUs), meeting SLOs on inference latency with the least amount of resources.

Despite all these efforts, virtually all inference serving systems suffer efficiency issues when applied to \emph{hybrid DL} due to the following reasons: 
(1) Dynamic DNN partitioning in hybrid DL produces \emph{non-uniform} DNN fragments that cannot be served with shared DNN instances on servers.
Therefore, a separate DNN instance may need to be provisioned for each individual mobile client with a small request rate, leading to limited request batching opportunities critical for resource efficiency.
Also, these DNN fragments change continuously due to network dynamics of the mobile clients.
(2) Existing systems mostly employ coarse-grained GPU allocation~\cite{nsdi17-clipper,usenix20-clockwork}, where DNNs are encapsulated in containers that occupy GPUs exclusively in a time-sharing manner~\cite{atc22-tetris}.
This leads to considerable GPU under-utilization~\cite{2019-sosp-nexus,atc22-tetris}.
Figure~\ref{fig:bg:underutilization} depicts the throughput achievable by the allocated resources versus the actual demand from two mobile clients each issuing 30 requests per second (RPS). 
With Clipper the allocated GPU resources can support more than an order of magnitude higher request rates, even with opportunistic sharing (a single instance is used when the two mobile clients have the same network condition and thus the same partition point) enabled between the two mobile clients.

\subsection{DNN Re-alignment and Challenges}

\begin{figure}
	\centering
	\includegraphics[page=2,width=0.42\textwidth]{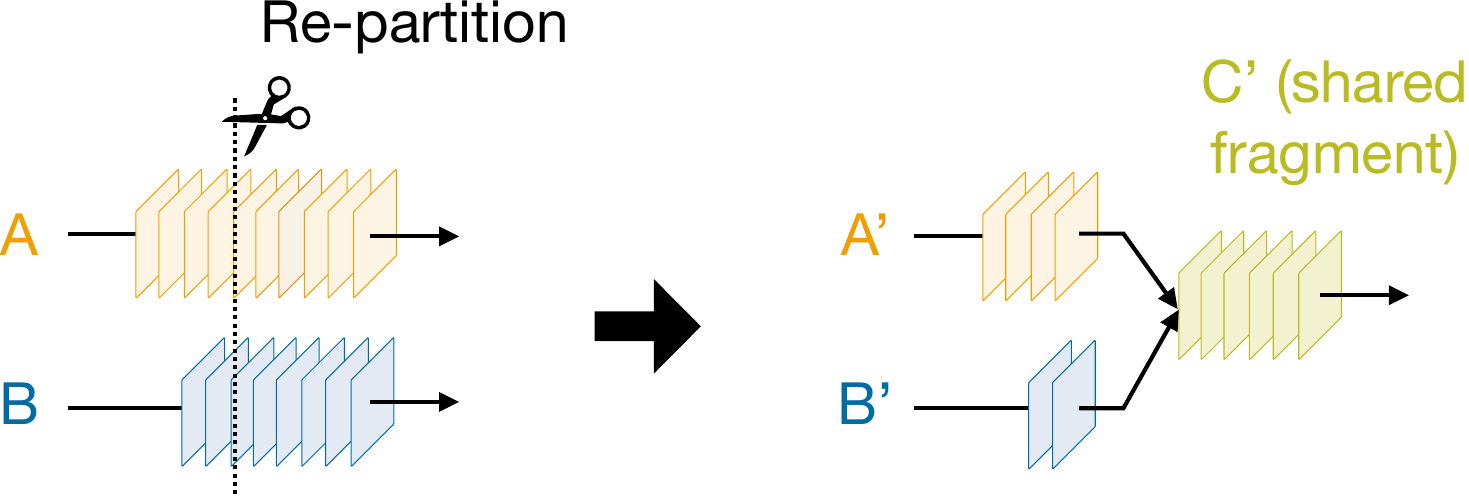}
	\caption{DNN re-alignment via re-partitioning.}
	\label{fig:bg:re-alignment}
\end{figure}

To address the non-uniformity issue, we observe that a simple technique called \emph{DNN re-alignment} could help. 
The key idea is to re-partition DNN fragments to generate aligned DNN fragments shared by multiple mobile clients, as depicted in Figure~\ref{fig:bg:re-alignment}.
This works since DNN fragments in hybrid DL, despite being misaligned, share their last layers. 
This sharing behavior can even be observed across different DNNs due to the widespread adoption of pre-trained foundation models and transfer learning (for fine-tuning the final layers) ~\cite{atc22-tetris,imc21-mobile-measurement}. 
Re-alignment leverages such sharing to increase the request batching opportunity, aiming to achieve better resource efficiency.

However, DNN re-alignment amplifies DNN fragmentation, which, when combined with coarse-grained GPU allocation, leads to poor GPU utilization. 
Fortunately, the state-of-the-art CUDA MPS (Multi-Process Service) allows for sharing a GPU spatially across multiple processes to improve utilization.
Based on MPS, recent work like GSLICE with fine-grained GPU allocation demonstrates significant resource savings for inference serving~\cite{socc20-gslice}.
Therefore, we adopt MPS and use GPU share (\%) to represent the amount of GPU resources.
%

Efficiently applying DNN re-alignment for inference serving in hybrid DL is non-trivial.
In particular, we identify the following three major challenges.

\begin{figure}[!t]
	\centering
	\subfloat[]{
		\includegraphics[scale=0.35]{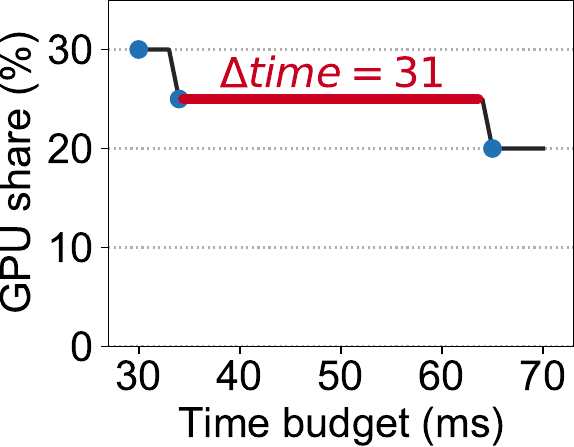}
		\label{fig:bg:resource-elasticity:latency}
	}
	\hfill
	\subfloat[]{
		\includegraphics[scale=0.35]{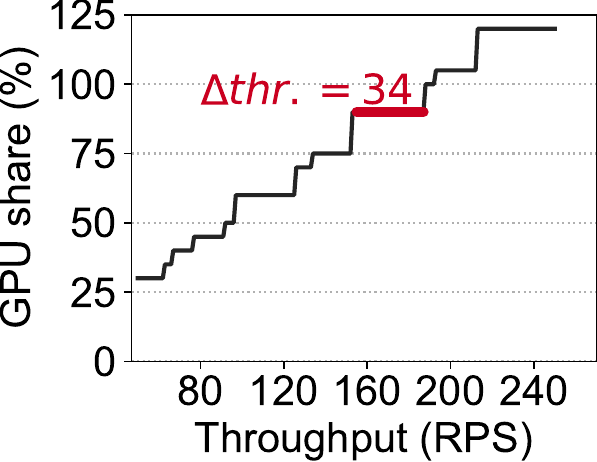}   
		\label{fig:bg:resource-elasticity:thr}
	}
	\caption{Discreteness in resource consumption of Inception-v3: (a) required GPU share to meet different inference time budgets with throughput of 200RPS, (b) required GPU share to achieve different inference throughput while maintaining inference latency of 25ms.}
		\label{fig:bg:resource-elasticity}
\end{figure}

\mypara{How to merge DNN fragments?}
There may exist DNN fragments that are uniform in structure, but serve requests with different time budgets and/or different request rates.
Merging the requests for these fragments and using one instance to serve them could potentially increase the batching opportunity, but the time budget of all requests will need to follow the smallest one among all merged requests. 
This hurts resource efficiency since some more relaxed time budget will be wasted.
Interestingly, we observe that smaller time budgets and higher throughput do not always translate into higher resource consumption, as shown in Figure~\ref{fig:bg:resource-elasticity}. 
This is caused by the discreteness of the batch size and the resource unit. 
This property can be leveraged to merge DNN fragments to improve resource efficiency.


\mypara{How to group DNN fragments for re-alignment?}
Given a set of non-uniform DNN fragments, deciding the subset of the DNN fragments to be grouped for re-alignment is critical. 
DNN fragments can be heterogeneous with respect to multiple factors: the partition point, the time budget, and the request rate. 
There are two decisions to make for fragment grouping, namely, how many DNN fragments to put in one group and which DNN fragments to put together?
Intuitively, a larger group size provides more opportunities for optimization, but comes with exponentially higher time complexity for making grouping decisions. 
The grouping decision should be made based on the compound influence of the heterogeneity factors on resource efficiency, which needs to be explored carefully.

\mypara{How to select the re-partition point?}
After grouping, we need to perform re-alignment by selecting a layer to re-partition the DNN fragments in the same group. 
This re-partition point can be at any layer between the partition point of the largest fragment and the last layer. 
Hence, deciding the re-partition point consuming minimum resources requires to explore all the possible layers efficiently.

\section{\namex System Design}
\label{sec:system}

We present \namex---a new system for serving DNNs fragments in hybrid DL. 
The goal of \namex is to minimize the overall server resource consumption while guaranteeing latency SLOs for mobile applications.

\begin{figure}[!t]
	\centering
	\includegraphics[width=0.45\textwidth,page=3]{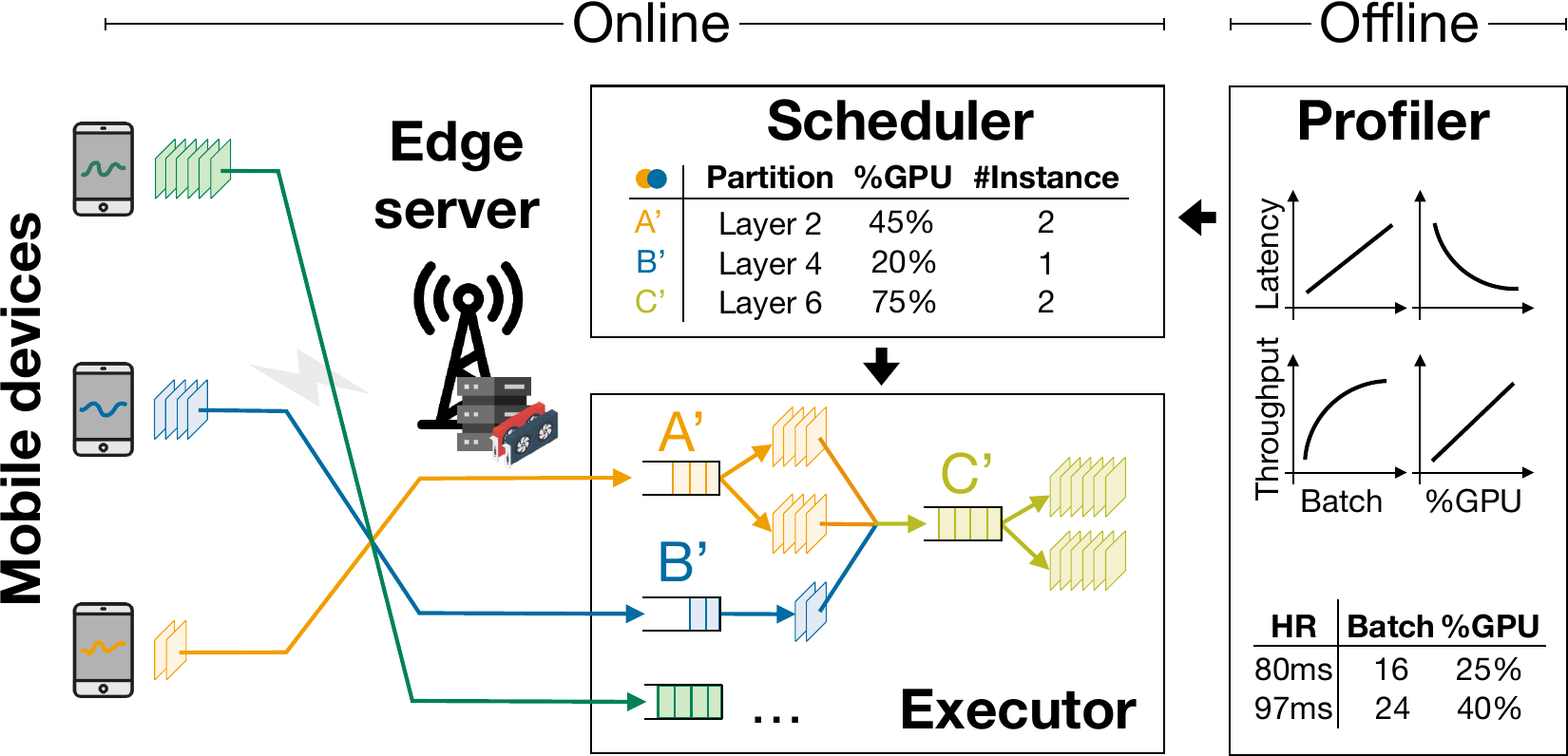}
	\caption{System architecture of \namex.}
	\label{fig:system:overview}
\end{figure}

\mypara{Overview.}
The system architecture of \namex is depicted in Figure~\ref{fig:system:overview}.
\namex consists of three core components: \emph{profiler}, \emph{scheduler}, and \emph{executor}, all running on the edge server.
The profiler runs offline while the scheduler and the executor run online at runtime.
We assume each mobile client applies hybrid DL by partitioning the DNN with any existing strategies and we focus on serving the DNN fragments with GPUs on the server side.
The general workflow of \namex is as follows: 
\jing{First, the profiler collects performance profiles of DNNs with respect to both latency and throughput in batch size and GPU share across all available GPUs in servers.
This ensures Graft is adaptable to heterogeneous GPU setups.}
Then, the scheduler takes the performance profiles and generates a resource-efficient execution plan. 
The execution plan consists of the DNN fragment groups, the re-partition point for each fragment in each group, the GPU share for each fragment, and the number of instances to spawn for each fragment.
Finally, the executor takes the execution plan and deploys the DNN fragment instances with specified GPU shares accordingly.
Here, the requests for each DNN fragment are evenly distributed to all its instances with a load balancer.
\jing{Additionally, the requests that fail to meet SLOs are dropped by load balancer for resource saving.}

\jing{
Due to effective optimizations (detailed in \S~\ref{sec:scheduling:allocating}), Graft maintains high time-efficiency in its decision-making.
This offers the opportunity to repeatedly invoke the scheduler when any mobile device changes its partition point due to significant network bandwidth changes.
This trigger-based approach adjusts its resource allocation timely based on real-time network conditions, significantly boosting the inference serving performance.}

\mypara{Scheduler.}
The scheduler provides the intelligence of the system by generating the execution plan, taking as input the offloaded DNN fragments, the request rate of each mobile client, and the performance profiles for the concerned DNNs from the profiler.
The goal of scheduling is to minimize the overall resource consumption for the given mobile clients, while guaranteeing their latency SLOs. 
To this end, the scheduler first applies an efficient algorithm to merge fragments (\S\ref{sec:scheduling:merging}), and then organizes DNN fragments into groups (\S\ref{sec:scheduling:grouping}).
Finally, it performs a fine-grained resource allocation algorithm with pruning-based optimizations (\S\ref{sec:scheduling:allocating}) to determine the most resource-efficient execution plan, which is then deployed by the executor to serve the mobile clients.

\section{Scheduling for DNN Re-alignment}
\label{sec:scheduling}

We now discuss the three steps the scheduler takes to generate the most resource-efficient execution plan, namely fragments merging, grouping, and re-partitioning.

\subsection{DNN Fragments Merging}
\label{sec:scheduling:merging}
Resource allocation for serving DNNs involves decision-making for multiple discrete variables including the number of DNN instances, unit of GPU share, and batch size. 
Such discreteness can lead to significant resource over-allocation when provisioning a large number of DNN fragments, i.e., a DNN instance can serve more requests without requiring more resources. 
We quantify resource over-allocation with a metric called \emph{\jing{resource margin}} defined as $(q_a-q_d)/q_d$, which captures the gap between the demanded throughput $q_d$ and the actual achievable throughput $q_a$ under given resources.

Reducing resource over-allocation is equivalent to minimizing the resource margin. 
To this end, \emph{\jing{we propose to merge incrementally uniform fragments that have the same partition point and time budget, until the resource margin decreases to a predefined threshold called the merging threshold.}}
Merging with such a threshold leaves more space for optimization in fragments grouping and re-partitioning, which turns out to be preferable for \namex over the strategy where we merge as much as possible. 
We will analyze the impact of the merging threshold in \S\ref{sec:exp:merging}.
Merging not only improves resource efficiency, but also reduces the number of fragments to deal with in the next steps of grouping and re-partitioning. 

\subsection{DNN Fragments Grouping}
\label{sec:scheduling:grouping}
The second step concerns the grouping of DNN fragments on the server into fragment subsets such that the total resource consumption by the DNN fragments is minimized.
Obtaining the optimal grouping of fragments is hard, due to the combinatorial explosion. 
One insight we can leverage is that \emph{\jing{it is beneficial to group fragments with similar properties (i.e., partition point, time budget, and request rate)}}. 
Applying this insight, we show that \emph{\jing{the grouping problem can be converted into a variant of the classic balanced graph partitioning problem}}~\cite{balanced-graph-partitioning},which aims to divide the vertices of a graph into $K$ equal-sized and disjoint subsets while minimizing/maximizing the total edge costs across different subsets.
Intuitively, we can construct a complete graph with all DNN fragments as nodes and for the edges between node pairs, we assign weights based on the similarity of the fragments represented by the nodes on each edge. 
Now, the grouping problem becomes dividing the nodes into $K$ equal-sized and disjoint subsets on the constructed graph. 
The goal is to maximize the total edge weights in all subsets, equivalent to minimizing the total edge weights across subsets.


Solving the balanced graph partition problem is still non-trivial~\cite{balanced-graph-partitioning}.
Existing work typically applies heuristics that either maximize internal edge costs within subsets or minimize external edge costs. 
Fennel et al. propose a framework that combines heuristics in both directions~\cite{wsdm14-fennel}.  
We follow a similar approach and define an objective function integrating the variance of the edge weights in each subset and the total edge weights across subsets, formulated as
\begin{equation}
\min\sum_{k=1}^K\sum_{e\in \mathbf{E}_k}\frac{(w_e-{\bar{w_k}})^2}{|\mathbf{E}_k|} + \sum_{k=1}^K \sum_{e\in \mathbf{E}'_k} w_e,
\label{alg:group:obj}
\end{equation}
where $w_e$ represents the weight on edge $e$, $\bar{w_k}$ is the average edge weight in subset $k$, $\mathbf{E}_k$ and $\mathbf{E}'_k$ are the sets of internal and external edges for subset $k$, respectively.
We assign the edge weight using the weighted Euclidean distance between the property vectors (consisting of partition point $p$, time budget $t$, and inference throughput $q$) of the fragments on each edge, i.e., $|\vec{v}_i, \vec{v}_j|$ where $\vec{v} = \left \langle p,t,q \right\rangle$. 


\minor{To strike the trade-off between resource efficiency and time complexity, we design a greedy algorithm following the approach in~\cite{wsdm14-fennel} with the above objective function}: 
(a) Randomly select $K$ fragments to initiate the groups. 
Here, $K$ is the number of expected groups and is dictated by the group size. 
Note that the group size does not constrain the batch size, which is decided in resource allocation during re-partitioning. 
(b) Iterate over the remaining fragments and assign each of them to the group with the least increase of cost given by Formula~(\ref{alg:group:obj}).
We will explore the impact of group size and factor weights in the Euclidean distance calculation in \S\ref{sec:exp:grouping}.

\subsection{DNN Fragments Re-partitioning}
\label{sec:scheduling:allocating}
We now focus on deciding the re-partition point and resource allocation for the fragments in each group, aiming to minimize the overall resource consumption with latency SLO guarantees.
\emph{\jing{Re-partitioning breaks down the fragment execution into two stages (see Figure~\ref{fig:bg:re-alignment}): the alignment stage where fragments have their own parts to execute concurrently, and the shared execution stage where the same instances are used for all fragments.
Given a re-partition point, deciding the resource allocation is equivalent to allocating appropriate time budget to each of the two stages, since the time budget dictates the required resources in each stage.}}

The re-alignment algorithm is listed in Algorithm~\ref{alg:re-alignment}.
Given a group $\mathbf{F}_G$ of fragments, we loop through all possible re-partition points (line 5) between $\min\{p_{1},\dots,p_{M}\}$ and $p_E$, where $p_{i}, i \in [1,M]$ denotes the partition point of fragment $f_i$ in the group and $p_E$ denotes the last layer of the DNN.
Under each re-partition point $p$, we can further divide the fragments into two parts: $\mathbf{F}_A$ that will be re-partitioned and $\mathbf{F}_B$ that will not be affected (line 7).
For $\mathbf{F}_A$ we apply the re-partitioning at layer $p$ and obtain all possible time budget allocation schemes denoted by $\mathbf{D}$, where each allocation scheme is given by $\mathbf{d} = \langle d_1,...,d_{|\mathbf{F}_A| + 1} \rangle$ and $d_i$ is the time allocated to fragment $f_i$.
Here, $d_{|\mathbf{F}_A| +1}$ represents the time allocated for the common fragment shared among all fragments in $\mathbf{F}_A$ generated by the re-partitioning. 
A valid time budget allocation scheme has to ensure that the total time spent by a request is lower than the available time budget of that request. 
\emph{Considering the worst-case queueing delay which equals the execution time~\cite{2019-sosp-nexus}, the total execution time of each fragment after re-partitioning has to be no larger than half of the total available time budget (line 8)}.
\minor{Meanwhile, batch sizes (denoted as $\mathbf{b}$) are self-adapted with respect to re-partition points and time budgets, rather than fixed settings.}
For each legitimate scheme $\mathbf{d} \in \mathbf{D}$, we calculate the minimum required resources and find out the scheme with the smallest resource consumption (lines 9-12). 
The above procedure will be recursively applied to the fragment set $\mathbf{F}_B$ until no fragment is left (line 13). 
The final step is to find the re-partition point and the corresponding min-resource execution plan that leads to the lowest overall resource consumption (line 16).

\begin{algorithm}[!t]
    {
	\caption{Fragments re-partitioning algorithm}
    \label{alg:re-alignment}
	\LinesNumbered 
	\KwIn{$\mathbf{F}_G=\{\left\langle p,t,q \right\rangle\}$: a group of fragments}
	\KwOut{$\left\langle p_{OPT}, X_{OPT} \right\rangle$: min-resource re-partition point and corresponding execution plan}
	\SetKwFunction{FMain}{\texttt{realign}}
    \SetKwProg{Fn}{Function}{:}{}
    
    \Fn{\FMain{$F$}}{
    
    \If{$\left| \mathbf{F} \right| = 0$} 
    {
     \textbf{return} $\emptyset$
     }
	 
	    $\mathbf{W} \leftarrow \emptyset$ \\
		\ForEach {$p \in [\min\{p_{1},\dots,p_{M}\},p_E]$}
		{
			$r_{min} \leftarrow \infty,~X \leftarrow \emptyset $ \\
			$\mathbf{F}_A \leftarrow \{f_i~|~f_i \in \mathbf{F} \wedge p_i < p\},~\mathbf{F}_B \leftarrow \mathbf{F} \setminus \mathbf{F}_A$\\
			
			$\mathbf{D} \leftarrow \{\langle d_1,\dots,d_{| \mathbf{F}_A|+1} \rangle~|~d_i+d_{\left| \mathbf{F}_A \right|+1} \leq \min\{t_{j}~|~f_j \in \mathbf{F}_A\}/2 \wedge i \in [1,|\mathbf{F}_A|] \}$ \\
			\ForEach {$\mathbf{d} \in \mathbf{D}$}
			{
			    $\left\langle \mathbf{r}, \mathbf{b}, \mathbf{c} \right\rangle \leftarrow$ \texttt{min\_resource}($p,\mathbf{d}$)\\
				\If{$\sum{\mathbf{r}} \le r_{min}$}
				{
					$r_{min} \leftarrow \sum{\mathbf{r}\circ \mathbf{c}}$,~
					$X \leftarrow \left\langle \mathbf{r}, \mathbf{b}, \mathbf{c} \right\rangle$
				}
			}
			
			$\mathbf{w} \leftarrow \{ \left\langle p, X \right\rangle \cup  $~\FMain{$\mathbf{F}_B$}\} \\
			\text{return}~$\mathbf{W} \cup \{\mathbf{w}\}$
		}
}
$\mathbf{S} \leftarrow$ \FMain{$F_G$} \\
$\left\langle p_{OPT}, X_{OPT} \right\rangle \leftarrow \arg\min_{ \mathbf{w} \in \mathbf{S}} \sum{\mathbf{r} \circ \mathbf{c}}$\\
		\textbf{return} $\left\langle p_{OPT}, X_{OPT} \right\rangle$
}
\end{algorithm}

The time complexity of re-alignment algorithm can be analyzed as follows.
The merging step uses ``mergesort'' to sort the fragments in $O(n\log n)$ time where $n$ is the total number of fragments.
The grouping step is transformed into a variant of balanced graph partitioning, which is solved by our proposed greedy method in $O(n^2)$ time.
The re-partitioning step loops through $O(n/M)$ groups, and in each group the re-partitioning problem (concerning $M$ fragments) can be solved by the Simplex algorithm in polynomial time~\cite{gurobi}. 
We will explore the time complexity of \namex empirically in \S\ref{sec:exp:overhead}. 

\namex applies several important optimizations to reduce its time overhead: (1) DNN fragments merging (\S\ref{sec:scheduling:merging}) to reduce the number of fragments, (2) graph building based on a simple metric and heuristics to group fragments with an adjustable group size (\S\ref{sec:scheduling:grouping}), and (3) pruning search space by reserving only the most ``efficient'' solutions (e.g., the blue dots in Figure~\ref{fig:bg:resource-elasticity:latency}) and parallelism for re-partitioning for different groups.

\section{Evaluation}
\label{sec:evaluation}
We have implemented a prototype of \namex and conducted extensive experiments to validate its performance.

\subsection{Setup and Implementation}
\setlength{\tabcolsep}{1pt}
\begin{table}[]
	\small
	\centering
	\caption{\label{table:exp:setup:mobile_specs} Specifications of mobile devices.}
	\vspace{-0.3cm}
	\begin{tabular}{@{}l  c c c c @{}}
		\toprule
		\textbf{Device} &  \textbf{GPU} & \textbf{AI perf.} & \textbf{Mem.} &  \textbf{Mode} \\
		\midrule
		Nano & 128-core Maxwell & 472 GFLOPS& 4G & MAXN~\cite{nano-energy-mode} \\
		TX2 & 256-core Pascal & 1.33 TFLOPS &8G & MAXQ~\cite{tx2-energy-mode} \\ 
		\bottomrule
	\end{tabular}
\end{table}

\minor{We assess the effectiveness of Graft across three different experimental scales, namely, small-, large- and massive scales, which are categorized based on the number of involved mobile devices.
In small-scale experiments, four NVIDIA Jetson Nano Developer Kits and two NVIDIA Jetson TX2 Developer Kits are used as mobile devices to evaluate Graft's performance under homogeneous/heterogeneous mobile scenarios.
The specifications of mobile devices are given in Table~\ref{table:exp:setup:mobile_specs}.
Following existing work \cite{mobicom20-spinn,arxiv18-jointdnn}, we choose Jetson boards since they have comparable specifications to real-world low- and high-end mobile devices, but are much simpler to program and control.
In large-scale experiments, we follow the widely adopted approach where we use two additional servers (in total 20 CPU cores) to emulate 20 mobile devices~\cite{socc20-gslice,mobicom20-spinn}.
Additionally, we also use simulation to evaluate \namex under massive mobile devices (e.g., at the level of thousands).
All servers and Jetson boards are connected to the same switch with 10GbE and 1GbE links, respectively.}

The data path of the system is implemented as follows. 
Each DNN fragment is materialized as multiple DNN instances, each of which works as an independent process, by DNN runtime Pytorch v1.9.0. 
The DNN is partitioned across the mobile device and the edge server based on Neurosurgeon~\cite{asplos17-neurosurgeon} for its simplicity; other partitioning strategies can be used as well (explained in \S\ref{sec:appendix}).
As Figure~\ref{fig:system:overview} shows, a request, once completes on-device processing, is first sent, via a network socket, to the server and then buffered as a tensor in a queue for the corresponding fragment.
This queue is shared by all the instances for each DNN fragment, which process requests in batch from the queue. 
At server-sides, \namex utilizes the Unix domain socket to transfer intermediate result (tensors) among DNN fragments.
\jing{Moreover, we set up a thread pool with a capacity of five for each DNN instance to efficiently pipeline their transferring.}

The control path implements the scheduling algorithms in Python. 
For implementing the merging and grouping algorithms, we leverage packages \texttt{numpy} and \texttt{networkx}.
For calculating the optimal re-partition point and resource allocation, i.e., execution plan, we use \texttt{cvxpy} with the \texttt{GUROBI} solver, which takes the profiles of DNN fragments with respect to both latency and throughput in batch size and GPU share.
Based on the plan, \namex starts DNN instances accordingly while terminating the outdated instances in an asynchronous manner. 
The network bandwidth between the mobile device and the server is shaped by replaying real-world 5G bandwidth traces~\cite{mmsys20-5g} (trace snippet shown in Figure~\ref{fig:bg:partition_point}), using the Linux \texttt{tc} utility (detailed in ~\S\ref{sec:appendix}).

We use five representative DNNs from TorchVision v0.10.0: four image classification models namely Inception-v3 (Inc), ResNet-101 (Res), VGG11 (VGG), and Vision Transformer-B16 (ViT), and a semantic image segmentation model called DeepLabV3 MobileNetv3-Large (Mob)~\cite{axiv17-mobilenet}.
The input size for the DNNs is around 588KB~\cite{socc20-gslice}.
\minor{DNNs, except for ViT, have request rates of 30RPS from each mobile device.
Due to its high mobile latency (column 6 in Table~\ref{table:exp:setup:mobile_info}), ViT has a lower request rate of 1RPS to prevent excessive frames blocking on mobile sides, which could deteriorate user experience.}
Table~\ref{table:exp:setup:mobile_info} lists the parameters and inference latency of these models.
The latency SLO for the considered models is typically lower than their mobile inference latency; otherwise, there is no need to offload to a server.
We set the SLO of each model to 95\% of the model's mobile inference latency. 
Later, we will explore the performance of \namex under varying SLOs in~\S\ref{sec:eval:sla}.

\setlength{\tabcolsep}{6pt}
\begin{table}[!t]
	\small
	\centering
	\caption{\label{table:exp:setup:mobile_info} Parameters and inference latency of used models when executing on mobile GPUs with batch size as one and executing on server GPUs with GPU share as thirty and batch sizes as one.}
	\vspace{-0.3cm}
	\begin{tabular}{@{}l  r r r r r @{}}
		\toprule
		\textbf{Model} &  \textbf{Inc} & \textbf{Res} & \textbf{VGG} &  \textbf{Mob} &  \textbf{ViT}\\
		\midrule
		Number of layers & 17 & 16 & 6 & 18 & 15\\
		Mobile latency [Nano] (ms) & 165 & 226 & 147 & 84 & 816\\
		Mobile latency [TX2] (ms) & 94 & 114 & 77 & 67 & 603\\ 
        Server latency (ms) & 29 & 30 & 6 & 19 & 58 \\
		\bottomrule
	\end{tabular}
	
\end{table}

The focus of \namex is on inference serving on the edge server.
Therefore, we use the state-of-the-art inference serving system GSLICE~\cite{socc20-gslice} and its enhanced version GSLICE$^+$, the optimal allocation (denoted as ``Optimal''), and the static allocation (Static) and its enhanced version (Static$^+$) as our baselines.
Note that Static and Static$^+$ depend on the average bandwidth of each mobile device to decide its resource allocation.
Neither GSLICE/GSLICE$^+$ nor Static/Static$^+$ conducts DNN re-alignment.

\begin{figure*}[!t]
	\centering
	\subfloat[Partition layer]{
		\includegraphics[width=0.9\textwidth]{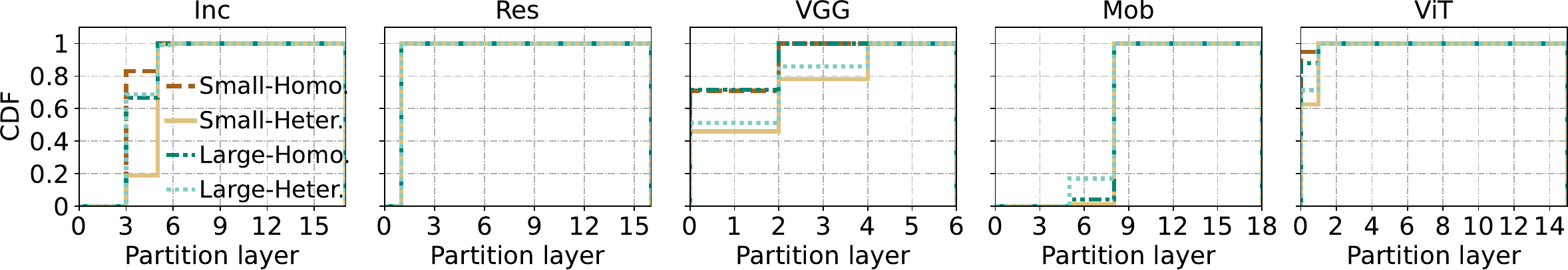}
		\label{fig:exp:overall:frag:partition-point}
	}
	\vfill
	\subfloat[Time budgets]{
		\includegraphics[width=0.9\textwidth]{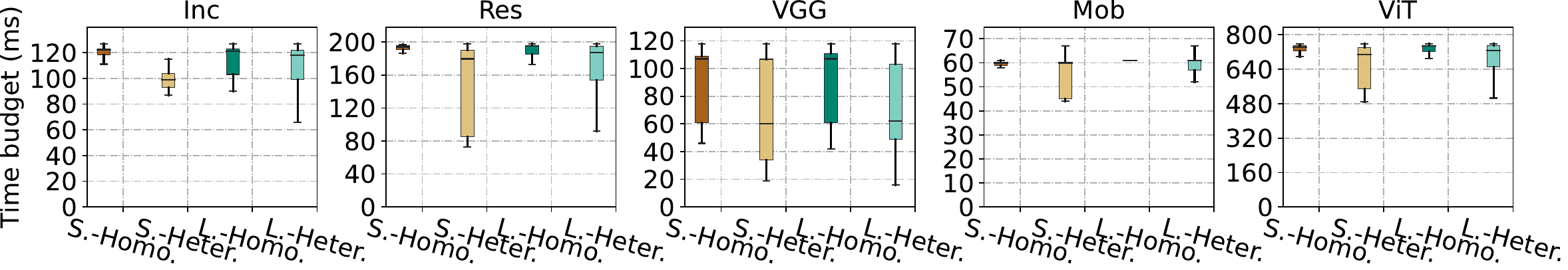}
		\label{fig:exp:overall:frag:budgets}
	}
 	\vspace{-0.2cm}
	\caption{Initial partition layer and time budgets of DNN fragment. S./L. represents small and large scales respectively.}
	\vspace{-0.2cm}
\end{figure*}

We evaluate two performance metrics namely the total resource consumption and the end-to-end latency (i.e., SLO guarantees).
Meanwhile, for a fair comparison, the baselines follow the same hybrid DL setup as \namex.
Specifically, requests' statistics, namely initial partition points and time budgets, over the small- and the large-scale scenarios, are depicted in Figure~\ref{fig:exp:overall:frag:partition-point} and Figure~\ref{fig:exp:overall:frag:budgets}, respectively.
\minor{
Note that Res, Mob, and ViT exhibit relatively polarised partitioning, with fragments displaying high similarities in initial partition points.
This is attributed to their unique architectures, which enables significant reductions in transmission loads at specific layers.
Taking Mob as an instance, layer one achieves reductions by 71.1\% compared with raw input.
Consequently, Neurosurgeon makes its polarised decisions to reap lower latency.}

The experiments involving the Optimal, namely \S5.2, \S5.3, \S\ref{sec:exp:grouping}, \S\ref{sec:exp:overhead} and \S\ref{sec:eval:sla}, are repeated by 10 times considering their considerable time cost;
meanwhile, the remaining experiments are conducted by 50 times, unless specified otherwise.
The resource allocation for each fragment on GPU is enforced via MPS with the resource unit of 1\% GPU share.
To reduce the interference among concurrent instances, we empirically cap the allocated GPU shares of each GPU to be lower than 100\%~\cite{socc20-gslice}.

\subsection{Performance in Small-Scale Scenarios}
\label{sec:exp:overall-performance}


\begin{table}[!t]
	\small
	\centering
	\caption{\label{table:exp:overall:resource} Overall resource reduction by \namex compared with GSLICE and GSLICE$^+$ under small- and large-scale, respectively.}
	\vspace{-0.3cm}
	\begin{tabular}{@{}l r r r r r r@{}}
		\toprule
		\multicolumn{2}{r}{\textbf{Model}} &  \textbf{Inc} & \textbf{Res} & \textbf{VGG} &  \textbf{Mob} &  \textbf{ViT}\\
		\midrule
		\multirow{2}{*}{Small-scale} & Homo. (\%) & 35.8 & 24.2 & 19.1 & 25 & 70.0 \\
		& Heter. (\%) & 24.2 & 16.3 & 31 & 26.4 & 63.1\\
		\midrule
		\multirow{2}{*}{Large-scale} & Homo. (\%) & 30.5 & 5.8* & 16.5 & 29.2 & 11.6* \\
		& Heter. (\%) & 20.4 & 16.1 & 30.3 & 41.1 & 59.6\\
		\bottomrule
	\end{tabular}\\
	* Case where GPU memory is the bottleneck.
\end{table}

\begin{figure*}[t]
	\centering
	\subfloat[Small-scale, homogeneous]{
		\includegraphics[scale=0.4]{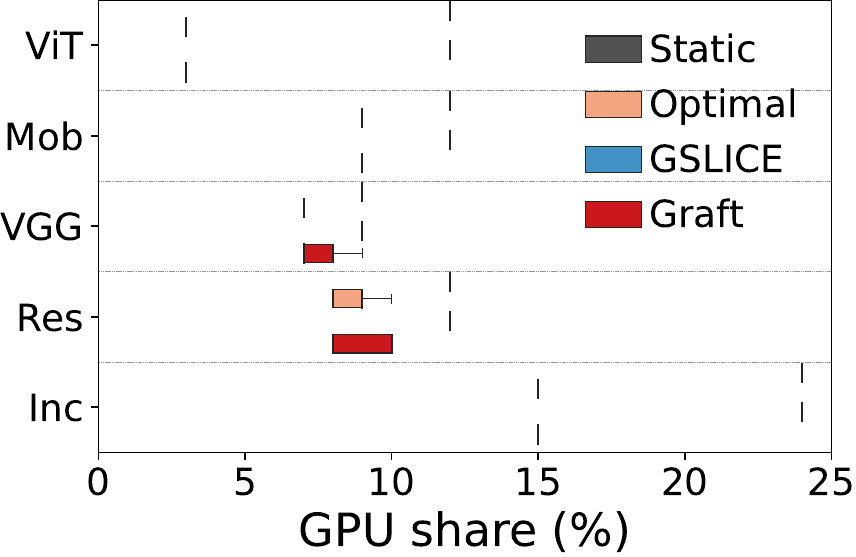}
		\label{fig:exp:overall:res:homo-small-scale}
	}
	\hfill
	\subfloat[Small-scale, heterogeneous]{
		\includegraphics[scale=0.4]{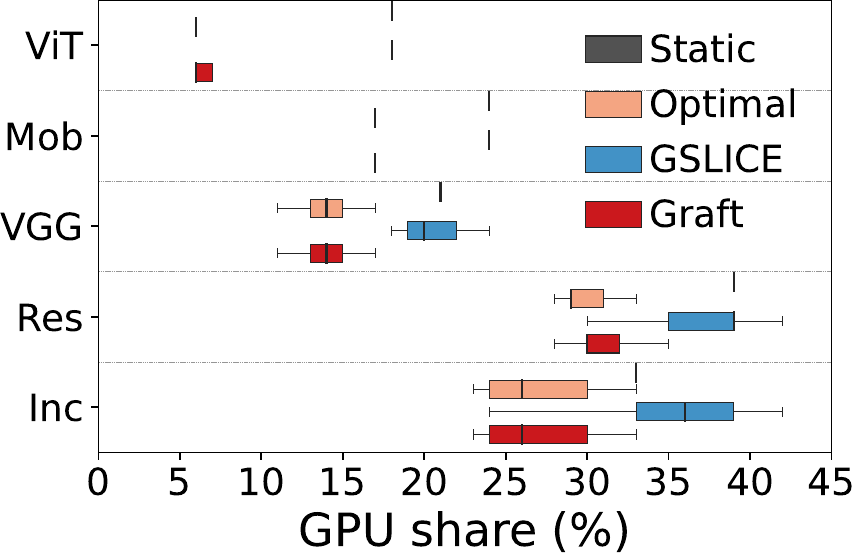}
		\label{fig:exp:overall:res:here-small-scale}
	}
	\hfill
	\subfloat[Large-scale, homogeneous]{
		\includegraphics[scale=0.4]{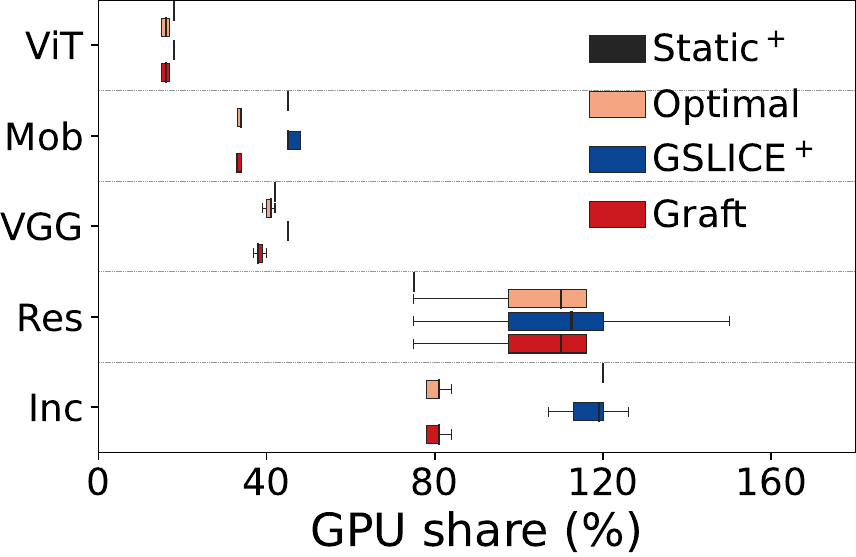}
		\label{fig:exp:overall:res:nano-large-scale}
	}
	\vspace{-0.2cm}
	\caption{Resource consumption comparison in (a) small-scale scenarios with four Nano boards, (b) small-scale scenarios with four Nano and two TX2 boards, and (c) large-scale scenarios with 20 emulated mobile clients.}
	\vspace{-0.2cm}
\end{figure*}


\mypara{Homogeneous cases.}
We compare the \emph{resource consumption} of \namex, GSLICE, Optimal, and Static when serving the models.
Here, we use four Jetson Nanos as mobile devices. 
Table~\ref{table:exp:overall:resource} (row 1) shows the average resource reduction of \namex when compared with GSLICE and Figure~\ref{fig:exp:overall:res:homo-small-scale} depicts the detailed comparison.
We can see that \emph{\namex outperforms GSLICE and Static significantly and performs close to Optimal}.
\emph{This improvement is attributed to DNN re-alignment, which allows for a larger batch size for each fragment, thus improving the resource efficiency}.

\begin{figure*}[!t]
	\centering
	\includegraphics[width=0.9\textwidth]{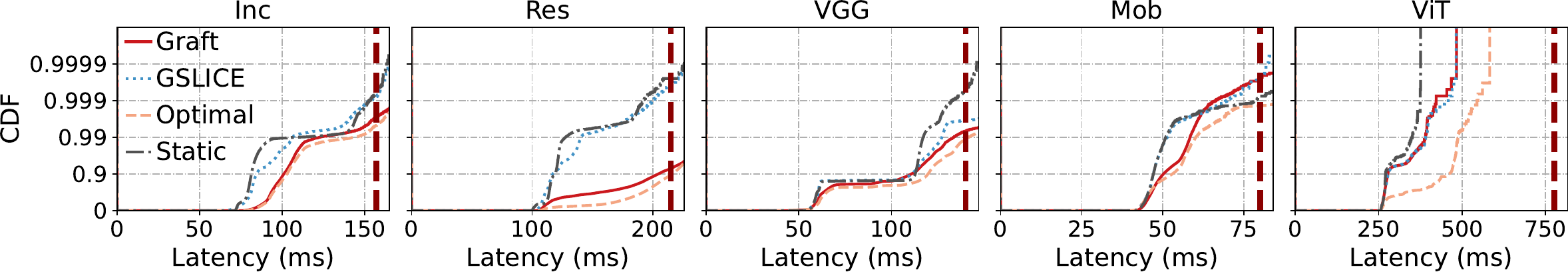}
	\vspace{-0.2cm}
	\caption{End-to-end latency distribution comparison in small-scale scenarios with four Nano boards. (The vertical dashed line marks the latency SLO).}
	\label{fig:exp:overall:cdf:homo-small-scale}
	\vspace{-0.3cm}
\end{figure*}

Figure~\ref{fig:exp:overall:cdf:homo-small-scale} shows the \emph{end-to-end latency} distribution.
We observe that both Static and GSLICE outperform \namex with Inc, Res, VGG, and ViT, due to its resource over-allocation.
Notably, \namex achieves lower latency with Mob.
This is because Mob has relatively low heterogeneity in both partition points and time budgets in the small-scale scenario (depicted in Figure~\ref{fig:exp:overall:frag:partition-point} and Figure~\ref{fig:exp:overall:frag:budgets}), allowing \namex to re-align the four fragments simultaneously.
Moreover, considering SLO guaranteeing, \namex depends on the lowest time budget of the fragments to make its decision, which speeds up the execution of some fragments than they demand, thus achieving lower latency. 

\mypara{Heterogeneous cases.}
We now compare the \emph{resource consumption} of \namex, GSLICE, Optimal, and Static under the heterogeneous scenarios, with four Jetson Nanos and two Jetson TX2s as mobile clients.
Table~\ref{table:exp:overall:resource} (row 2) shows the average resource reduction of \namex when compared with GSLICE; details are given in
Figure~\ref{fig:exp:overall:res:here-small-scale}.
Meanwhile, \namex achieves resource reduction by 13.9\%, 21\%, 33.9\%, 28.8\%, and 63.1\% for Inc, Res, VGG, Mob, and ViT, respectively when compared with Static.
We can observe from Figure~\ref{fig:exp:overall:frag:budgets} that Jetson TX2 generates fragments with significantly lower time budgets than those from Jetson Nanos;
meanwhile, the difference in partitioning points of DNN fragments also magnifies as illustrated in Figure~\ref{fig:exp:overall:frag:partition-point}. 
Facing this heterogeneous scenario, \namex tends to re-align the fragments separately based on their mobile devices, for maintaining high resource efficiency without violating SLOs.
This prevents \namex from building large batches across heterogeneous devices, especially for DNNs with limited resource margins (e.g., Res), thus affecting its effectiveness.
On the contrary, large gains are observed with VGG.
This is attributed to its high resource margins, which accommodate variations in time budgets and allow for larger batches that span across heterogeneous devices.

\begin{figure*}[t]
	\centering
	\subfloat[Jetson Nano]{
		\includegraphics[width=0.9\textwidth]{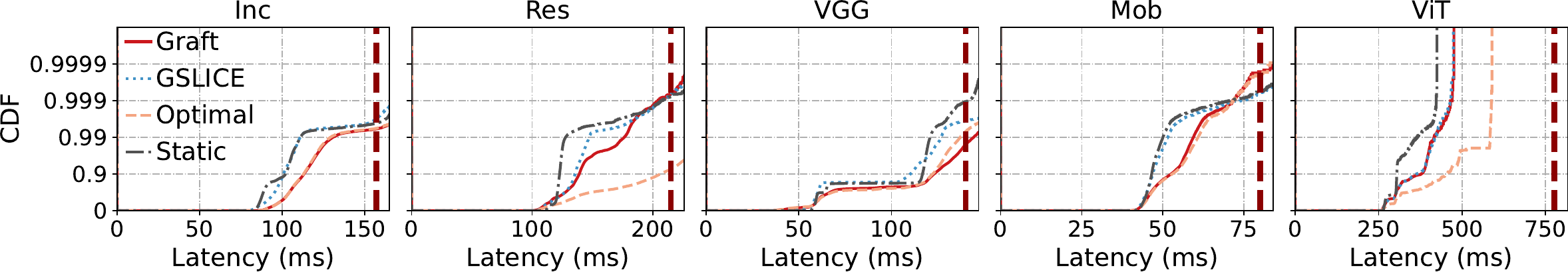}
		\label{fig:exp:overall:cdf:hete-nano-small-scale}
	}
    \vfill
	\subfloat[Jetson TX2]{
		\includegraphics[width=0.9\textwidth]{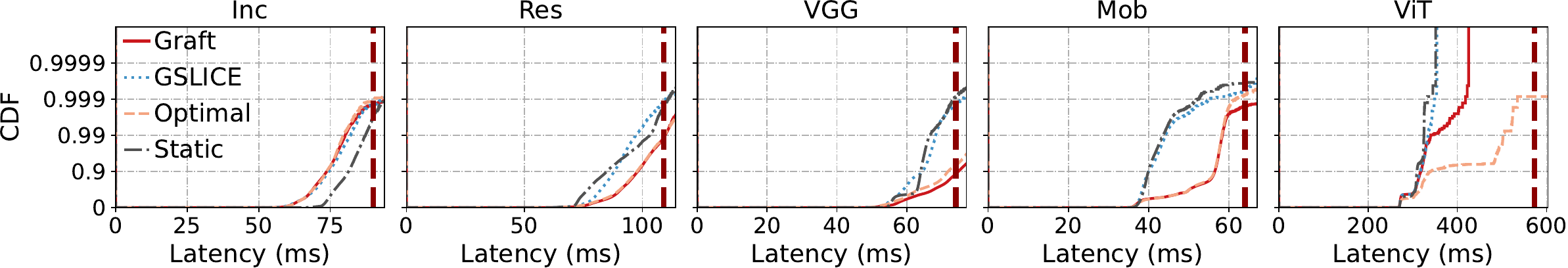}
		\label{fig:exp:overall:cdf:hete-tx2-small-scale}
	}
	\vspace{-0.3cm}
	\caption{End-to-end latency distribution comparison in small-scale heterogeneous scenarios for (a) Nano and (b) TX2 boards.}
	\vspace{-0.4cm}
\end{figure*}
\begin{figure*}[!t]
	\centering
	\includegraphics[width=0.9\textwidth]{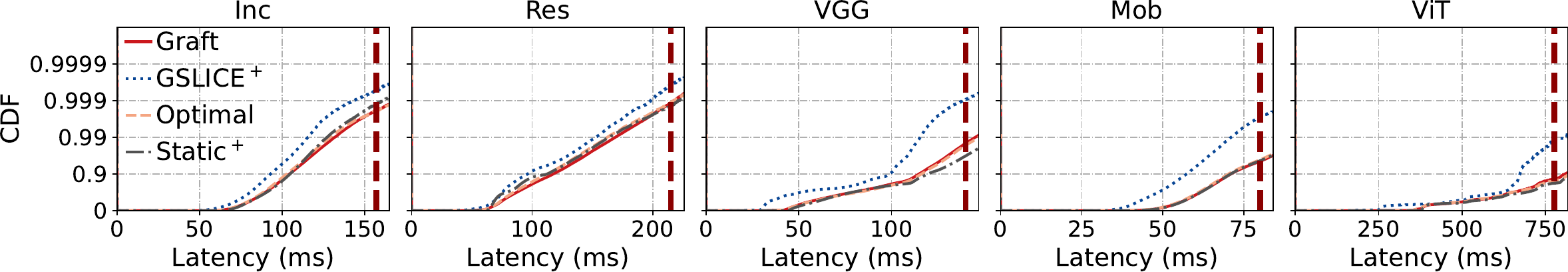}
	\vspace{-0.1cm}
	\caption{End-to-end latency distribution comparison in large-scale scenarios with 20 emulated mobile clients. (The vertical dashed line marks the latency SLO).}
	\label{fig:exp:overall:cdf:nano-large-scale}
	\vspace{-0.3cm}
\end{figure*}

We depict the \emph{end-to-end latency} distribution of models deployed on Nano and TX2 in Figure~\ref{fig:exp:overall:cdf:hete-nano-small-scale} and Figure~\ref{fig:exp:overall:cdf:hete-tx2-small-scale}, respectively.
We observe that \namex meets latency SLOs in most cases, except the case with VGG on Jetson TX2.
The reason is that VGG re-aligns the fragments from the heterogeneous mobile devices, thus forming large batches. 
While improving the resource efficiency, the increased batch size leads to the variation in queueing delay and incurs long-tail latency, especially for TX2 which has tight SLOs.

\subsection{Performance in Large-Scale Scenarios}
\mypara{Homogeneous cases.}
We compare the \emph{resource consumption} of \namex, GSLICE, Optimal, and Static under 20 mobile devices (emulated by 20 CPU cores).
The increased number of mobile devices introduces uniform fragments that share the same initial partition point and time budget.
For a fair comparison, we enable the best merging strategy (i.e., merge all uniform fragments) for GSLICE and Static and denote the enhanced version as GSLICE$^+$ and Static$^+$.
To prevent GPU memory overflow, we set an upper bound for the number of instances to five for each fragment based on empirical observations.
This bound is removed in large-scale simulations in \S\ref{sec:exp:large-scale}.

Table~\ref{table:exp:overall:resource} (row 3) presents the average resource reduction of \namex when compared with GSLICE$^+$; details are in Figure~\ref{fig:exp:overall:res:nano-large-scale}.
We can see that \emph{\namex consumes significantly fewer resources than GSLICE$^+$, and the gap between \namex and Optimal is within 4\%}.
Notably, the constraints in the number of instances for each fragment narrow down the optimization space, degrading the performance of \namex.
In particular, Res and ViT are bottlenecked by their significant requirements in GPU memory, beyond our testbed's capacity, and get diminished improvements in resource efficiency. 
This constraint also leads to resource over-allocation for Res as 63\% and 37.1\% of \namex than GSLICE and Static$^+$, respectively.
Overall, \emph{\namex reduces the resource consumption by 36.1\%, 20.8\%, 46.9\%, and 74.7\% for Inc, VGG, Mob, and ViT, respectively, when compared with GSLICE.
The reduction is 32.9\%, 9.1\%, 25.7\%, and 11.6\% as compared with Static$^+$}.

Figure~\ref{fig:exp:overall:cdf:nano-large-scale} depicts the \emph{end-to-end latency} distribution where \namex provides SLO guaranteeing for the five DNNs.
Note that GSLICE$^+$ provides lower end-to-end latency due to its considerable resource over-allocation when compared with \namex and Optimal.

\mypara{Heterogeneous cases.}
We also evaluate \namex in large-scale heterogeneous scenarios, where we emulate 15 Nano and 5 TX2 boards as mobile devices. 
Applying the algorithms of \namex, we calculate the resource consumption and compare it with that of GSLICE$^+$.
\minor{As shown in Table~\ref{table:exp:overall:resource} (row 4), even when provisioning DNNs with polarised partitioning, namely Res, Mob, and ViT (explained in \S5.1), which are particularly suitable for GSLICE$^+$ to merge and batch due to their similar initial partition points, \namex still achieves lower resource consumption by 16.1\%, 41.1\%, and 59.6\%, respectively.
This is attributed to \namex's comprehensive exploration in re-partition points and fine-grained re-allocation of time budgets (detailed in \S\ref{sec:scheduling:allocating}), which constructs substantial optimization spaces for improving resource efficiency.}

We were not able to obtain the end-to-end latency distribution due to the lack of GPU memory to support that number of DNN fragments under strict SLOs. 

\subsection{Effectiveness of Re-partitioning}

\begin{figure}[!h]
	\centering
	\includegraphics[width=0.4\textwidth]{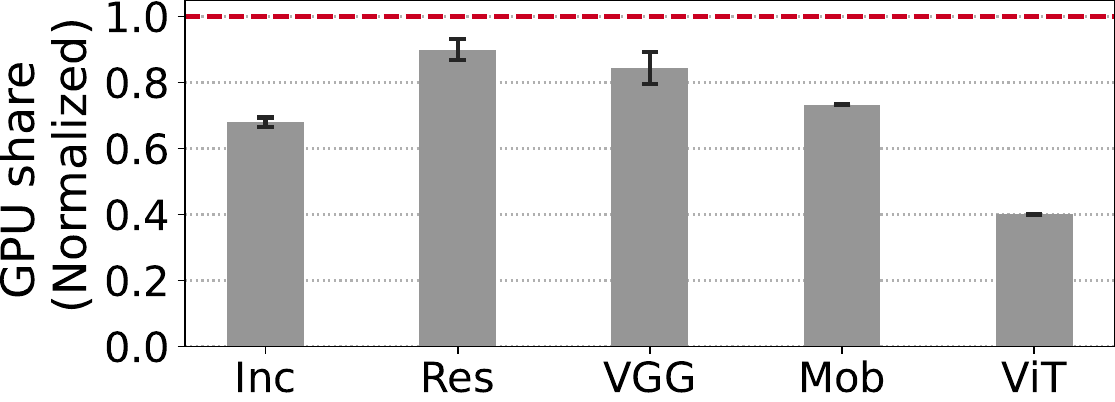}
	\vspace{-0.3cm}
	\caption{Resource consumption with re-partitioning, normalized by the case without re-partitioning.}
	\label{fig:exp:re-align}
	\vspace{-0.5cm}
\end{figure}

We now evaluate the resource consumption with and without re-partitioning when provisioning five random fragments, irrespective of the impact of merging and grouping.
Here, each fragment has the same frame rate as in \S\ref{sec:exp:overall-performance} and replays a random bandwidth from the real-world network trace.
Figure~\ref{fig:exp:re-align} shows that \emph{re-alignment reduces the resource consumption by up to 32\%, 10\%, 15.6\%, 26.7\% and 60\% for Inc, Res, VGG, Mob, and ViT, respectively}.
The improvement for Res is smaller due to its relatively lower resource margin.

\begin{figure}[!h]
	\centering
	\subfloat[]{
		\includegraphics[scale=0.35]{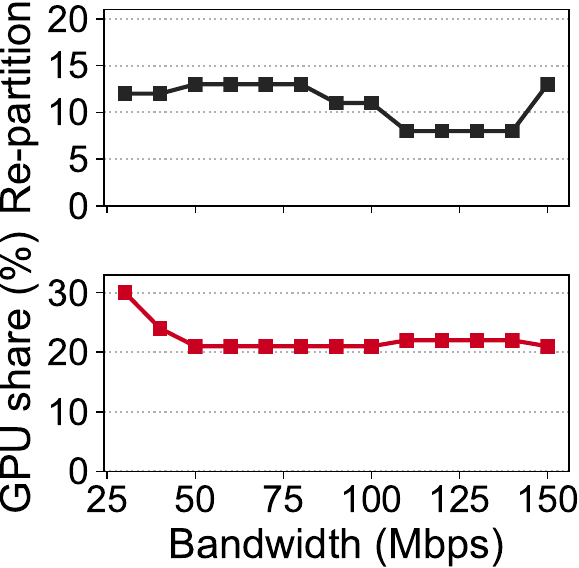}
		\label{fig:exp:scheduling:bw}
	}
	\subfloat[]{
		\includegraphics[scale=0.35]{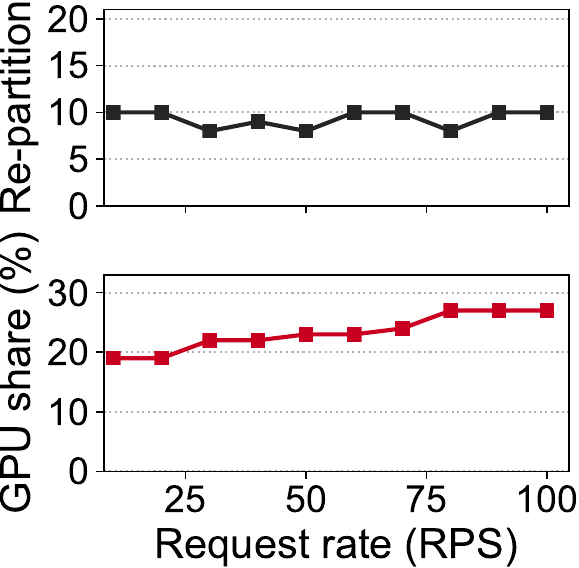}
		\label{fig:exp:scheduling:thr}
	}
	\caption{Re-partition point and GPU share of Inc under varying (a) network bandwidth and (b) request rate.}
	\vspace{-0.3cm}
\end{figure}

We also analyze how well re-partitioning performs under different network bandwidths and request rates.
To this end, we evaluate the re-partition point and the required GPU share while keeping four fragments unchanged and varying the configuration for the fifth fragment.
Taking Inc as an example, we observe from Figure~\ref{fig:exp:scheduling:bw} that \emph{the resource consumption decreases marginally with the increase of the network bandwidth}.
The reason can be seen in Algorithm~\ref{alg:re-alignment} that re-alignment is dictated by the lowest time budget of fragments in the same group.
With the increase of network bandwidth, the partition point of the fragments is fixed at layer three. 
However, their time budget difference keeps increasing until the time budget of the static fragments becomes the bottleneck for re-alignment, resulting in a marginal decrease in resource consumption.
Additionally, we observe in Figure~\ref{fig:exp:scheduling:thr} that \emph{high request rates lead to higher resource consumption.
As for the re-partition point, it is influenced by the DNN type, partition point, time budget, and request rate, simultaneously}.

\subsection{Effectiveness of Merging}
\label{sec:exp:merging}
\begin{figure}[!h]
	\centering
	\includegraphics[width=0.42\textwidth]{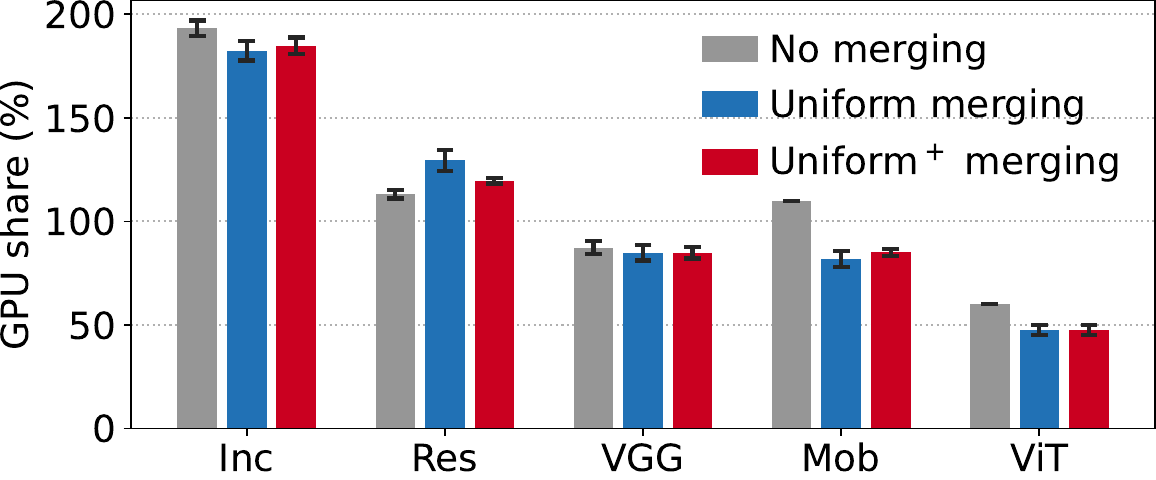}
	\caption{Resource consumption with different merging strategies (50 fragments with merging threshold 0.2).}
	\label{fig:exp:merging:overall}
	\vspace{-0.2cm}
\end{figure}

\begin{figure}[!h]
	\centering
	\includegraphics[width=0.45\textwidth]{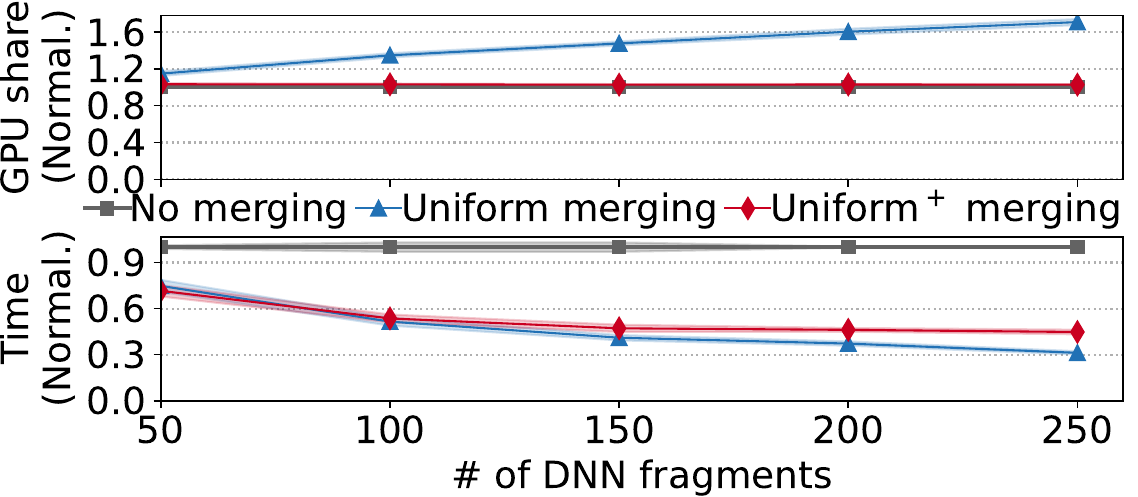}
	\caption{Resource consumption (top) and time cost (bottom) of Res normalized by no merging under different numbers of DNN fragments with merging threshold of 0.2.}
	\label{fig:exp:merging:resnet}
	\vspace{-0.2cm}
\end{figure}

We evaluate the resource consumption of \namex when employing three different merging strategies: No-merging, Uniform merging (i.e., merging all uniform fragments), and Uniform$^+$ merging (i.e., merging until reaching the merging threshold), respectively.
We generate 50 fragments and set the merging threshold as 0.2.
Figure~\ref{fig:exp:merging:overall} shows that, \emph{compared with Uniform$^+$ merging, Uniform merging suffers resource over-allocation when applied to Res}.
This occurs because Res fragments exhibit relatively low heterogeneity as shown in Figure~\ref{fig:exp:overall:frag:partition-point}, which leads to a number of fragments with high throughput when conducting Uniform merging. 
Moreover, Rec has constrained resource margins due to its high computation intensity.
The conflict, between high throughput and low resource margin, often leads to the independent execution of DNN fragments with no further realignment, which hurts the resource efficiency consequently.
In comparison, \emph{Uniform$^+$ merging limits the number of fragments involved in merging and controls the fragments throughout.
This creates reasonable space for realignment across fragments, thus achieving higher resource efficiency.
Furthermore, this improvement grows with the increase of fragment numbers} as shown in Figure~\ref{fig:exp:merging:resnet}(top).

It is worth noting that \textit{insufficient merging, e.g., adopting No merging or setting Uniform$^+$ with high merging threshold, also hurts resource efficiency}.
This is because it results in limited available batch sizes for realignment.
On the other hand, \emph{Uniform$^+$ merging greatly decreases the problem sizes for the following grouping and re-partitioning}.
We can see from Figure~\ref{fig:exp:merging:resnet}(bottom) that Uniform$^+$ merging reduces the number of DNN fragments by up to 42.6\%, 42.3\%, 50.7\%, 76.1\%, and 38.9\% on average for Inc, Res, VGG, Mob, and ViT respectively, as compared with No-merging.

\begin{figure*}[]
\centering
 	\subfloat[]{
		\includegraphics[scale=0.36]{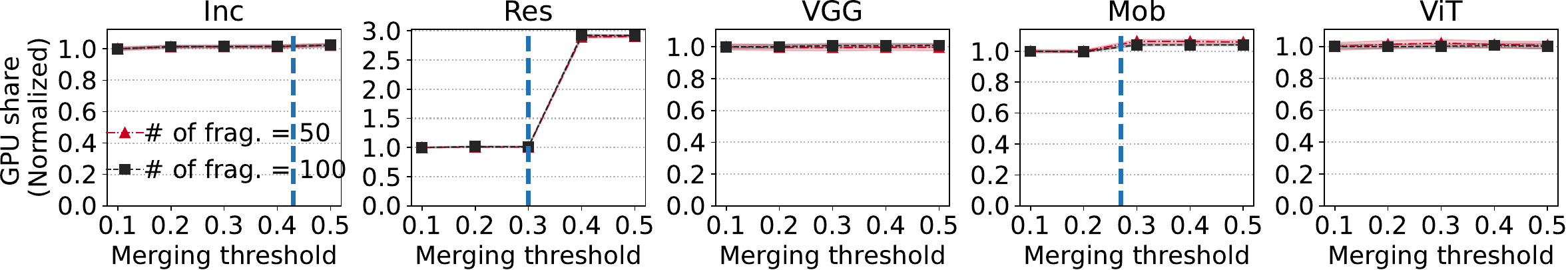}
		\label{fig:exp:merging_thresh}
	}
	\subfloat[]{
		\includegraphics[scale=0.35]{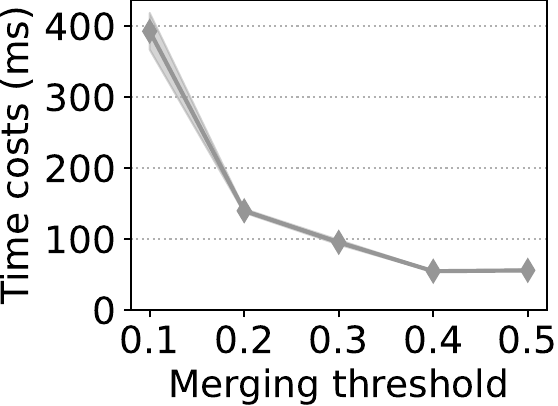}
		\label{fig:exp:merging_overhead}
	}
 \vspace{-0.2cm}
	\caption{(a) Resource consumption under varying merging thresholds and varying numbers of DNN fragments, normalized by that under merging threshold of 0.1. (The vertical dashed line is the resource margin, which for VGG and ViT is 0.58 and 3, respectively.) (b) Time costs of Uniform$^+$ merging when operating 25 fragments of Res under varying thresholds.}
	\vspace{-0.2cm}
\end{figure*}

We analyze the sensitivity of Uniform$^+$ merging to different merging thresholds under varying fragment numbers.
As illustrated in Figure~\ref{fig:exp:merging_thresh}, \emph{most DNNs except Res are insensitive to the variation of merging thresholds}.
This is thanks to the interaction between merging and the subsequent grouping and re-partitioning.
Specifically, when the merging threshold increases, the number of fragments involved in merging decreases, limiting the fragments throughput consequently.
This enables grouping and re-partitioning to unlock the potential of re-alignment, i.e., grouping more fragments and building larger batches.
Otherwise, as the threshold decreases, the generated fragments are facilitated with larger batch sizes as well as higher resource efficiency, which, however, constrains the performance of grouping and re-partitioning due to SLO requirements.

\minor{Notably, based on our empirical experience, \textit{setting merging thresholds less than DNNs' resource margin helps \namex maintain the trade-off between time efficiency and resource efficiency}.
The reason is that this lower value compels \namex to sufficiently merge uniform fragments while retaining reasonable optimization space for grouping and re-partitioning.
This is in contrast to Uniform, which neglects preservation.
Take Res, which has a resource margin of 0.3, an instance.
As shown in Figure~\ref{fig:exp:merging_thresh}, Uniform$^+$ facilitates \namex with resource reduction by 2.9$\times$ when its merging threshold decreases from 0.4 to 0.2.
Meanwhile, it decreases the number of fragments by 14.5\% on average for the subsequent two steps.
However, as depicted in Figure~\ref{fig:exp:merging_overhead}, an excessive decrease in the threshold may incur considerable time costs during merging due to its incremental exploration (detailed in \S\ref{sec:scheduling:merging}).
Empirically, we set the merging threshold at 0.2 by default considering its satisfactory performance in striking the aforementioned trade-off.
}

\subsection{Effectiveness of Grouping}
\label{sec:exp:grouping}

We compare the resource consumption of \namex when adopting the proposed similarity-based grouping and the optimal grouping, respectively, to provision 25 fragments.
The proposed grouping utilizes the Euclidean distance as the edge weight to build the graph, instead of the exact amount of resources required by Optimal.
The result shows that the \emph{similarity-based grouping achieves comparable performance as more than 0.7\% compared with Optimal}.
Additionally, the similarity-based grouping ensures low overhead for \namex as detailed in~\S\ref{sec:exp:overhead}.

\begin{figure}[]
	\subfloat[]{
		\includegraphics[scale=0.4]{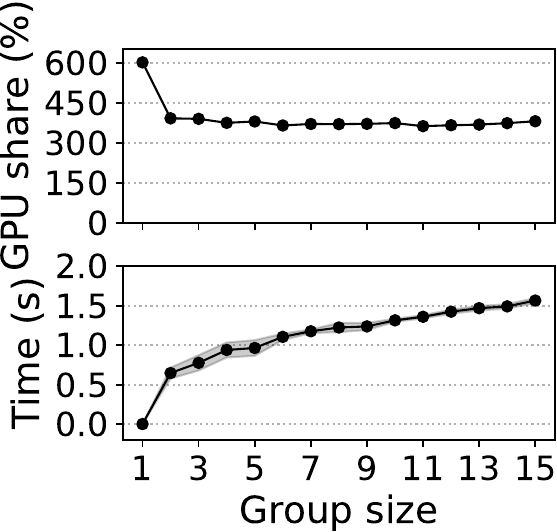}
		\label{fig:exp:gropuing:group_size}
	}
	\subfloat[]{
		\includegraphics[scale=0.4]{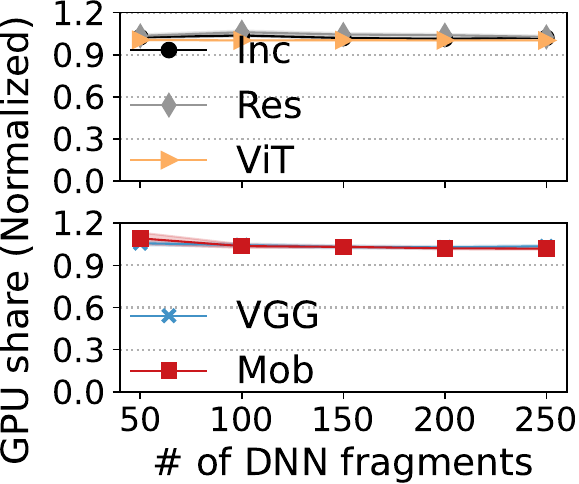}
		\label{fig:exp:gropuing:group_weight}
	}
    \vspace{-0.2cm}
	\caption{(a) Resource consumption and time cost of \namex provisioning Inc under varying group sizes; (b) Resource consumption adopting equal grouping weight normalized by the optimal weight.}
	\vspace{-0.3cm}
\end{figure}

We also analyze the impact of the group size and factor weights (namely for partition point, time budget, and throughput as detailed in~\S\ref{sec:scheduling:grouping}).
Taking Inc as an example, we observe from Figure~\ref{fig:exp:gropuing:group_size} that \emph{increased group sizes make a marginal improvement in resource consumption (top), but bring higher time cost (bottom).
Empirically, a group size of five is a good trade-off point between time complexity and resource savings}.

Figure~\ref{fig:exp:gropuing:group_weight} compares the resource consumption of \namex under equal factor weights and optimal factor weights.
We can see that the gap is no more than 4.1\% on average.
This means the \emph{optimization space for factor weights is quite limited}.
Based on our empirical experience, slightly increasing the weight for time budget for models with limited resource margins such as Res leads to higher resource efficiency, due to the smaller diversity of the fragments within each group.

\subsection{Achievable Throughput}
\begin{figure*}[t]
    \centering
	\includegraphics[width=0.9\textwidth]{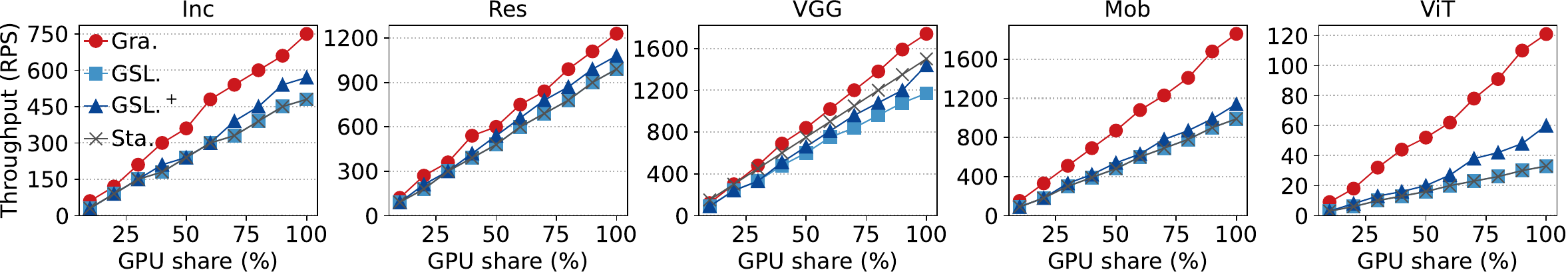}
	\caption{Achievable throughput of \namex, GSLICE, GSLICE$^+$, and Static under varying amounts of resources.}
	\label{fig:exp:thr}
	\vspace{-0.3cm}
\end{figure*}

We compare the maximum achievable throughput of \namex, GSLICE, GSLICE$^+$, and Static under restricted resource offerings.
In this experiment, we gradually increase the number of fragments until their required resources exceed an upper bound.
Figure~\ref{fig:exp:thr} shows that \emph{\namex achieves higher throughput as 1.45$\times$, 1.18$\times$, 1.29$\times$, 1.65$\times$, and 2.38$\times$ on average for Inc, Res, VGG, Mob, and ViT respectively when compared with GSLICE$^+$}.
This is because GSLICE$^+$ only conducts optimization (batching) among uniform DNN fragments, and ignores that across heterogeneous fragments, thus receiving limited improvement.
On the other hand, \emph{\namex achieves higher throughput as 1.57$\times$, 1.29$\times$, 1.4$\times$, 1.76$\times$, and 3.3$\times$ on average for Inc, Res, VGG, Mob, and ViT, respectively than GSLICE.
The improvement is as 1.57$\times$, 1.29$\times$, 1.09$\times$, 1.79$\times$, and 3.32$\times$, respectively, as compared with Static}.

\subsection{Massive-Scale Simulations}
\label{sec:exp:large-scale}
\begin{figure*}[t]
    \centering
	\includegraphics[width=0.9\textwidth]{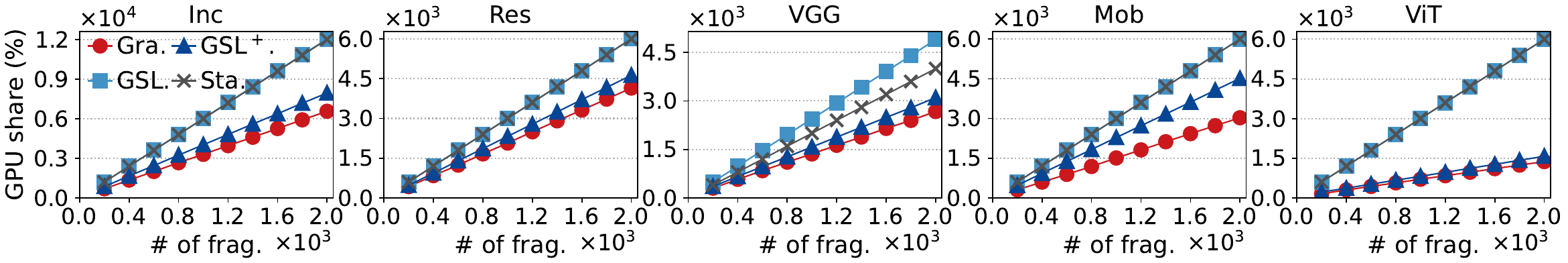}
	\caption{Resource consumption of \namex, GSLICE, GSLICE$^+$, and Static under large numbers of DNN fragments.}
	\label{fig:exp:large_scale}
	\vspace{-0.4cm}
\end{figure*}

We further evaluate the resource consumption of \namex, GSLICE, GSLICE$^+$, and Static with simulations. 
We focus on large-scale scenarios with thousands of fragments, which cannot be handled by our testbed.
For high time efficiency, we set the merging threshold as 0.01 for \namex.
Figure~\ref{fig:exp:large_scale} shows that \namex outperforms the baselines. 
Specifically, \emph{GSLICE and Static have approximate resource consumption, which 1.8$\times$, 1.44$\times$, 1.76$\times$, 2$\times$, and 4$\times$ on average for Inc, Res, VGG, Mob, and ViT, when compared with \namex.
Meanwhile, \namex reduces the resource consumption by 18.1\%, 10.6\%, 13.6\%, 34.1\%, and 16.1\% on average, compared with GSLICE$^+$, for the five models, respectively.
This means that \namex still outperforms the pure-batching based allocation (GSLICE$^+$) at scale}. 
Note that the low merging threshold incurs excessive merging for DNNs of low resource margins or of low heterogeneity (e.g., Res and ViT), compromising \namex's resource efficiency (discussed in~\S\ref{sec:exp:merging}).

\subsection{System Overhead}
\label{sec:exp:overhead}
\begin{figure}[h]
	\centering
	\subfloat[]{
		\includegraphics[scale=0.4]{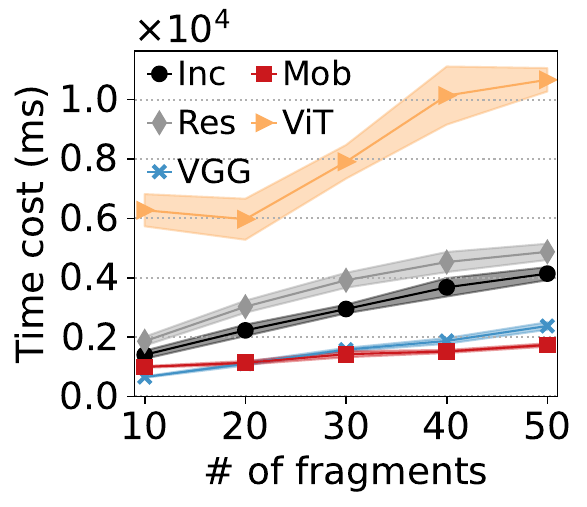}
		\label{fig:exp:overhead}
	}
	\subfloat[]{
		\includegraphics[scale=0.4]{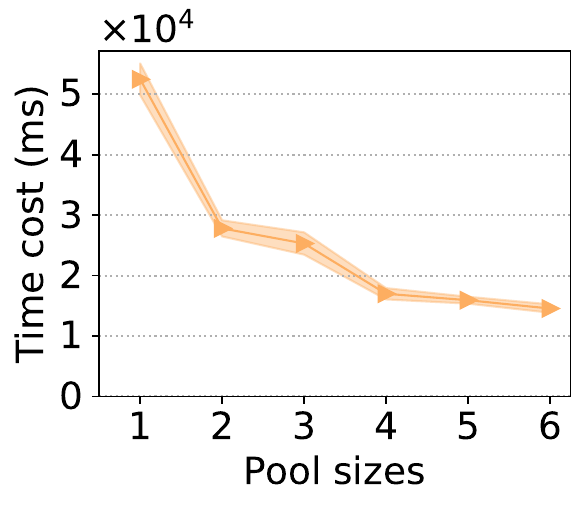}
		\label{fig:exp:speedup}
	}
    \vspace{-0.2cm}
	\caption{(a) Time cost of \namex when realigning fragments ranging from ten to fifty; (b) Time costs of \namex when realigning fifty fragments of ViT with pool sizes ranging from one to six.}
	\vspace{-0.3cm}
\end{figure}

We evaluate the time cost and memory footprint of \namex and Optimal respectively when re-aligning fragments ranging from ten to fifty.
Due to limited space, we only illustrate the time costs of \namex.
The time cost of \namex achieves an average reduction by up to 99.6\% across all five models, in comparison with Optimal.
The reason is that Optimal enumerates all the feasible groupings (e.g., 252 for Inc with ten fragments) to figure out the minimum resource consumption, incurring significant overheads.
In contrast, \namex only explores one grouping heuristically and thus guarantees time efficiency while achieving close-to-optimum resource efficiency. 

On the other hand, as shown in Figure~\ref{fig:exp:overhead}, \textit{the time cost of Graft increases as the number of fragments grows}.
This is because under given group sizes, a higher number of fragment yields more groups.
While each grouped fragment can be realigned independently, more groups inevitably incur higher time costs.
In particular, ViT experiences significant high time costs.
This is attributed to its highly heterogeneous time budgets, which hinder the effectiveness of merging and result in more groups.
To fix this issue, we can introduce a process pool to realign the grouped fragments concurrently.
Figure~\ref{fig:exp:speedup} illustrates the time costs when realigning fifty fragments of ViT, which generate six groups after merging.
We can see that the time costs decrease by 1.9$\times$ as the pool size increases from one to two.
Notably, the increase in the pool size has diminishing returns.
To guarantee time efficiency, we set the pool size as two by default.

In addition, we also evaluate the memory footprint of \namex during the above process.
\emph{The average memory consumption is negligible as 35.69MB, 32.14MB, 29.9MB, 34.4MB, and 25.3MB for Inc, Res, VGG, Mob, and ViT respectively.
Moreover, this footprint increases no more than 0.63MB with the number of fragments increasing from 10 to 100}.

\subsection{Impact of SLOs}
\label{sec:eval:sla}
\begin{figure}[h]
    \centering
\includegraphics[width=0.45\textwidth]{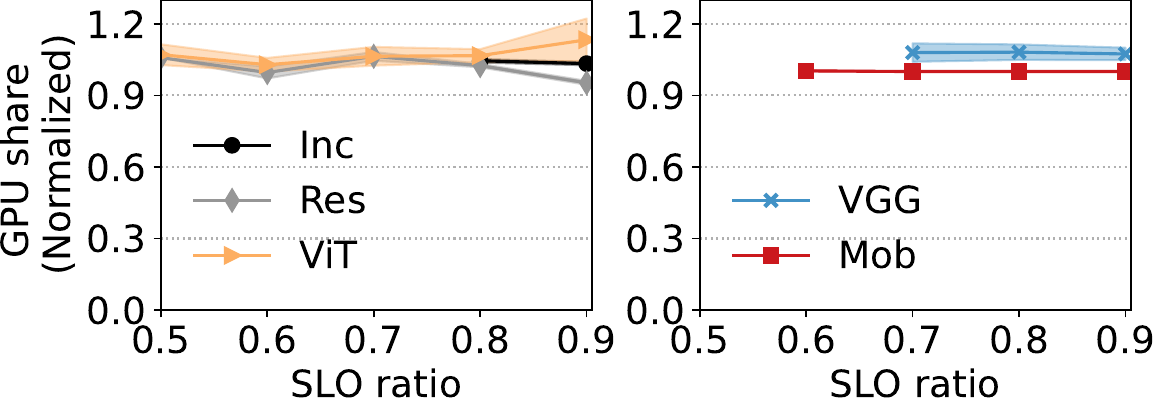}
	\caption{Resource consumption of \namex, normalized by Optimal, under varying SLO ratios}
	\label{fig:exp:slo}
	\vspace{-0.4cm}
\end{figure}

We evaluate the impact of SLOs on the performance of \namex. 
In our work, SLOs express DNN models' requirements in end-to-end latency.
For ease of exposition, the SLO for each model is specified as a ratio to its individual mobile inference latency and we call this ratio as SLO ratio, which can span from 0.5 to 0.9.
Moreover, the more strict performance requirements DNNs have in latency, the lower SLO ratios they should be set.
We take an e-commerce application as an example.
This application integrates multiple DNN models, each with specific performance requirements.
It is worth noting that \emph{these requirements vary depending on the business domains}~\cite{asplos23-erms}.
Specifically, for tasks deemed mission-critical, such as facial recognition for payment purposes, it is imperative to configure the corresponding DNN with relatively lower SLO ratios, to meet its stringent latency requirements. 
In comparison, tasks related to facial beautification, which exhibit a greater degree of tolerance to latency, can be set with higher SLO ratios.

Here, for the generality, we completely evaluate the impacts of SLO ratio over a broad spectrum, spanning from 0.5 to 0.9.
Note that hybrid DL (i.e., Neurosurgeon) may fail to find a feasible partition point when the SLO ratio is too small, e.g., less than 0.7 for Inc.
Figure~\ref{fig:exp:slo} shows that \emph{\namex is insensitive to the change of SLO (ratio), constantly achieving performance approximately as good as Optimal}.

\subsection{Performance in energy}
\label{sec:exp:energy}

\begin{figure}[!h]
	\centering
	\subfloat[Small-scale]{
		\includegraphics[scale=0.4]{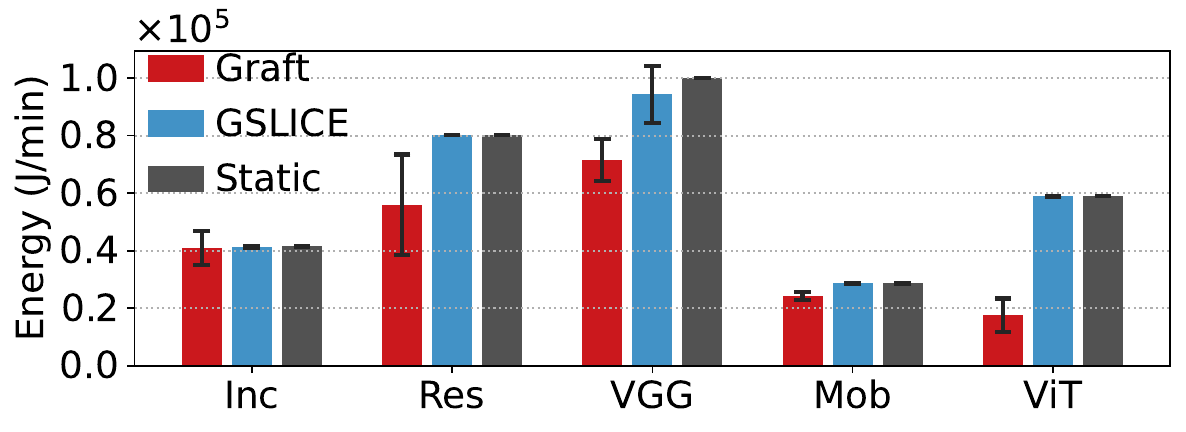}
		\label{fig:exp:energy:small-scale}
	}
	\vfill
	\vspace{-0.2cm}
	\subfloat[Large-scale]{
		\includegraphics[scale=0.4]{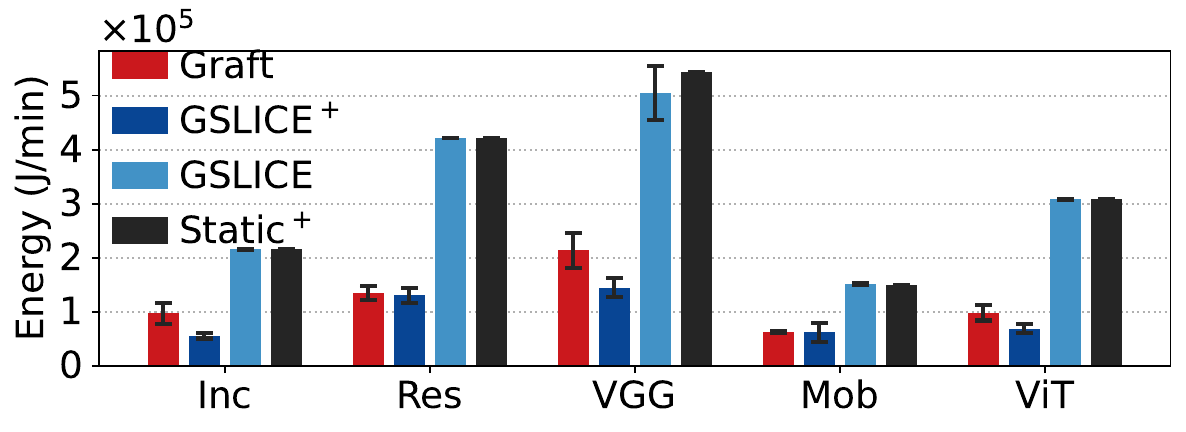}
		\label{fig:exp:energy:large_scale}
	}
	\vspace{-0.2cm}
	\caption{Energy consumption of \namex, GSLICE, GSLICE$^+$, Static, and Static$^+$ when realigning (a) four fragments and (b) twenty fragments.}
	\vspace{-0.3cm}
\end{figure}

\minor{
We evaluate the energy consumption of Graft, GSLICE, GSLICE$^+$, Static, and Static$^+$ under homogeneous small-scale and large-scale scenarios (with the same settings as in Section~5.2 and Section~5.3), respectively.
As depicted in Figure~\ref{fig:exp:energy:small-scale}, \textit{thanks to its higher resource efficiency, Graft achieves energy reduction by up to 70.1\% when compared with GSLICE and Static in the small-scale case}.
This improvement becomes more significant in the large-scale scenario due to the incorporation of a greater number of fragments in realignment.
Note that GSLICE$^+$ exhibits lower energy consumption than Graft by 42.7\%, 32.2\%, and 30.1\% on average for Inc, VGG, and ViT, respectively.
This is because of its sufficient merging enabled by Uniform (detailed in \S\ref{sec:exp:merging}), which enables GSLICE$^+$ with the opportunity to build larger batch sizes and thus reaps higher energy efficiency.
}

\section{\jing{Limitation and future work}}
\label{sec:limitation}
\mypara{Realignment disruption.}
Notably, there may exist cases where the realignment is disrupted by the DNN fragments that undergo frequent variations.
As a solution, we could introduce a shadow instances-based alternative as done in GSLICE in the future.
Specifically, this strategy sets up shadow instances for the latest arrived DNN fragments when the scheduler is busy with re-aligning existing fragments.
Upon completion, the scheduler identifies ``similar" fragments, which share the same partition points and approximate time budgets with the recently arrived ones, and then reuses their realignment on the latter.
This approach is motivated by the fact that smaller time budgets and higher throughput do not always dictate higher resource consumption due to the discreteness (e.g., batch size) in resource provisioning.
Meanwhile, partition points usually occur at several specific layers (illustrated in Figure~\ref{fig:exp:overall:frag:partition-point}), presenting a high potential for realignment reuse.
The above improvement could allow Graft to effectively handle the frequent variations of DNN fragments.

\mypara{Intermediate representations with diverse precision.}
In most cases, Graft does not require specific improvements to deal with intermediate representations' differences in precision.
This is because while intermediate representations may employ low bit-widths for lower transmission latency, most of them can be converted to full precision upon reaching servers~\cite{tecs22-dyno}, thus having no impacts on the subsequent realignment.
On the other hand, in cases where the conversion fails to proceed as expected, we can implement the following approach.
That is to separate DNN fragments based on their precision while restricting the realignment among the fragments with the same precision.

\mypara{Availability to other models.}
Recently early-exiting models are proposed to realize the trade-off between inference accuracy and inference latency~\cite{mobicom20-spinn,twc19-edgent,mm21-smtm}.
These models incorporate multiple exits, allowing requests to terminate at different stages of processing.
Moreover, prior to execution, requests lack precise information about their exact exiting layer.
In particular, when requests exit before reaching the re-partitioning point, the re-aligned DNN fragment will fail to construct predefined batch sizes and thus incur resource over-allocation.
\minor{
To address this issue, we can monitor the achievable throughput (e.g., requests per second) of the exiting layers that are located after the re-partitioning point.
Then, Graft adjusts batch sizes accordingly to reap higher resource efficiency}.

\minor{BERT, LSTM, RNN, and Transformer are designed to analyze sequential data.
Compared with CNN, these models have higher data dependency due to their specific requirements of ``memory''.
This consequently needs to facilitate Graft with an adequate data compression technique to control data exchanging overheads.
Notably, to the best of our knowledge, BERT, LSTM, RNN, and Transformer are rarely used alone but rather serve as basic building blocks to construct more complex architectures.
Moreover, for the sake of higher accuracy, modern models tend to integrate considerable numbers of blocks, for example, ViT consists of twelve Transformer based blocks.
This offers the opportunity to focus on the realignment of inter-blocks, rather than intra-blocks, so as to strike the trade-off between resource efficiency and time complexity.
As shown in Table~3, Graft reaps resource reduction by up to 70\% for ViT as compared with GSLICE/GSLICE$^+$, which only uses batching without realignment.}

\mypara{Heterogeneous models.}
At present, the realignment is exclusively applied to fragments that originate from the same DNN or, at most, only have slight difference in final layers due to transfer learning.
We leave more fine-grained per-layer re-alignment as future work.
\minor{To support complex scenarios that involves multiple different DNNs, we can introduce a straightforward separation strategy for fragments depending on their DNN types, and then apply Graft}.

\mypara{Split training.} 
Graft demonstrates its effectiveness in provisioning inference tasks on the server side.
To support split training, which incorporates both clients and servers to train models collaboratively~\cite{corr19-split-learning}, several considerations must be addressed due to their distinct characteristics compared with inference tasks: 
(1) Looser time budgets, (2) Higher GPU memory requirements~\cite{arxiv19-salus,atc21-zico}, and (3)
Larger transmission loads, which encompass batched intermediate data and generated gradients across clients and servers~\cite{www23-edgemove}.
As a result, Graft needs to adjust its objective.
Instead of focusing on minimizing GPU computing resources without violating SLOs, Graft should now prioritize minimizing GPU memory consumption while shortening training duration as much as possible.
To achieve this goal, the profiler should collect the profiling data with respect to training throughput and latency in batch sizes and memory consumption.
Simultaneously, it is necessary to integrate a series of transmission/storage-targeted optimizations, such as quantization and compression~\cite{mobicom20-spinn,sosp21-casync}, for reaping lower data transmission latency.

\mypara{Compatible with other hardware.}
Graft effectively enables fine-grained resource allocation on hardware that supports spatial sharing, e.g., CPUs and GPUs.
To seamlessly support temporal-sharing accelerators, Graft requires upgrading its executor, ensuring that re-aligned DNN fragments are executed in order.

\mypara{Distributed edge setups.}
Graft exploits DNN fragments re-alignment within one edge node.
To extend to distributed edge setups, it needs to facilitate a straightforward deployment strategy--first bin-packing-based deployment, which is widely adopted by commercial platforms like AWS Lambda and Alibaba Compute Function~\cite{socc22-owl,atc18-peek-bench}.
Specifically, this strategy selects the first available server that can accommodate the requirements of both the separate and the re-aligned fragments (shown in Figure~4) appropriately.
Consequently, it effectively minimizes time-consuming data transmission across servers while mitigating the risk of overwhelming server resources.

\section{Related Work}
\label{sec:relatedwork}
We summarize related work in hybrid DL, DL inference serving, and GPU utilization improvement in this section.

\mypara{Hybrid DL.}
Recently, much research has been done on hybrid DL by both academia and industry. 
Neurosurgeon~\cite{asplos17-neurosurgeon} first introduces DNN partitioning to decrease the end-to-end latency by exploiting the computation capability of a mobile device and cloud servers under dynamic network bandwidth. 
JointDNN~\cite{arxiv18-jointdnn} extends the idea to multiple-part partitioning and explores the properties of partition points for different DNN types. 
Neurosurgeon and JointDNN both focus on DNNs with linear architectures (e.g., AlexNet~\cite{nips12-alexnet} and VGG~\cite{iclr15-vgg}).
DADS~\cite{infocom19-dads} explores partitioning for DNNs based on directed acyclic graph (DAG) (e.g., GoogleNet~\cite{cvpr15-googlnet}) and proposes two DNN partitioning strategies for light and heavy network conditions, respectively. 
There are also studies on specific problems involving hybrid DL such as deploying DNNs at the edge~\cite{sec19-couper,socc18-ionn,tpds20-multi-task}, mobile handover~\cite{sec19-deepsave}, and privacy protection~\cite{asplos20-shredder}.
Early-exit has also been merged into DNN partitioning recently for further decreasing the inference latency and resource consumption~\cite{mobicom20-spinn,twc19-edgent,mm21-smtm}. 
Auto-split has been deployed in a commercial cloud to enable edge-cloud  collaborative DNN inference~\cite{sigkdd21-autosplit}.

\mypara{DL inference serving.}
A large amount of work has been done to improve DL inference serving, aiming at resource efficiency and latency SLO guarantee.
The industry develops systems, such as TensorFlow Serving~\cite{tensorflow-serving} and TorchServe~\cite{torchserve}.
Meanwhile, the academic proposes a variety of inference serving systems like Clipper~\cite{nsdi17-clipper}, Nexus~\cite{2019-sosp-nexus}, AsyFunc~\cite{socc23-asyfunc}, iGniter~\cite{tpds23-igniter}, Opara~\cite{ccfsys23-opara}, etc.
Clockwork exploits performance predictability and proposes a principled bottom-up approach to achieve low tail latency for inference serving~\cite{usenix20-clockwork}.
Systems like ALERT~\cite{usenix20-alter} and INFaaS~\cite{2021-atc-infaas} employ model adaptation to balance performance, accuracy, and energy efficiency at runtime.
Apart from single models, more complex model graphs have been considered~\cite{socc20-inferline,eurosys19-grandslam,2021-arxiv-llama}.
Different from the above coarse-grain allocation frameworks working at the whole GPU level, GSLICE~\cite{socc20-gslice} spatially shares GPUs among located DNNs and hence reaps high throughput with latency guarantee.
Besides single DNN inference, there are also some studies on the inference serving ``chains'' or ``graphs" that coordinate multiple DNNs, including GrandSLAM~\cite{eurosys19-grandslam} and InferLine~\cite{socc20-inferline}.
Powerchief~\cite{isca17-powercheif} focuses on mitigating the latency of multi-stage applications on power-constrained CMP by means of identifying and accelerating the bottleneck service while considering the power budget.

\mypara{GPU utilization improvement.}
Numerous spatial sharing based approaches have been proposed to improve GPU utilization.
Prophet~\cite{asplos17-Prophet} depends on interference models with respect to shared resources, to realize the ``safe'' co-location for latency-sensitive applications without violating SLO.
Yeung et al.~\cite{hotcloud20-gpu-prediction} propose a prediction engine to foresee the GPU utilization of heterogeneous DL workloads, to increase resource utilization while reducing performance interference between workloads.
FGPUs~\cite{rtas19-fgpus} introduces both compute and memory bandwidth isolation mechanisms to promise the predictability of the concurrent applications on GPUs.
Slate~\cite{ipdps19-slate} conducts a workload-aware scheduling for the concurrent kernels to improve the GPU utilization while minimizing resource contention.
Maestro~\cite{asplos17-GPUMaestro} explores the GPU resource partitioning at the granularity of both the streaming multi-process (SM) and the simultaneous multi-kernel to improve the system performance.
Apart from spatial sharing, temporal sharing and spatial-temporal sharing are also explored to improve GPU utilization~\cite{middleware18-olympian,ppopp17-effisha,arxiv19-salus,atc22-share-gpu}.

In summary, current work in hybrid DL, such as Neurosurgeon and JointDNN, focuses on partitioning a single DNN to optimize latency or power across one mobile device and one (edge) server, without considering the efficient serving of non-uniform DNN fragments on the server.
Meanwhile, existing DL inference serving systems only pay attention to full-size DNNs in the cloud, ignoring the resource efficiency issue in hybrid DL where DNNs are fragmented.
In contrast, \namex is optimized for hybrid DL and addresses the inefficiency in provisioning non-uniform fragments, based on a new concept called re-alignment.
\namex is orthogonal to GPU spatial- and temporal-sharing improvements, which can be further leveraged to improve the performance of \namex.

\section{Conclusion}
\label{sec:conclusion}

In this paper, we identified the DNN fragment misalignment problem in inference serving for hybrid DL and proposed a new concept called re-alignment to address it by promoting request batching and sharing.
We proposed \namex---a first-of-its-kind inference serving system for hybrid DL adopting re-alignment. 
We identified the challenges in building \namex and proposed efficient algorithms for fine-grained resource allocation 
for \namex.
Experiments based on a system prototype show that \namex achieves significant resource savings while providing latency SLO guarantee.


\section{Appendix}
\label{sec:appendix}

\mypara{Bandwidth trace replay.}
We use the Linux \texttt{tc} utility to replay real-world 5G bandwidth trace between mobile devices and the server.
Specifically, tc offers various queueing disciplines, abbr. qdisc, on clients' network interfaces to shape their traffic.
We select Hierarchy Token Bucket-based queueing discipline, i.e., HTB qdisc, considering its reliable capacity to ensure specific upper limits on bandwidth~\cite{tc-classful-qdiscs}.
To replay varying 5G network conditions, we encapsulate the control in a script, and execute it periodically (each minute by default as~\cite{atc19-edgewise}) on each mobile device, which regulates the bandwidth of devices' network interface to predefined values.

\mypara{Other DNN partition strategies.}
Due to simplicity, we leverage Neurosurgeon to construct the hybrid DLs context.
Specifically, Neurosurgeon concentrates on fundamental factors, namely network conditions, SLO requirements, and mobile device computing capacity, to decide one particular partition point.
Other influential factors, such as performance interference across simultaneous workloads~\cite{micro20-autoscale} and DVFS settings~\cite{ipsn23-dvfo} on mobile devices are also investigated. 
On the other hand, there are works that enable multiple partition points~\cite{arxiv18-jointdnn,tsc22-webar}.
This offers the potential to achieve lower end-to-end latency but at the cost of higher resource consumption on the server-side~\cite{tecs22-dyno}.
The reason is that multiple partition points dictate extra transmission latency, which compresses the time budgets for inference tasks on the server side.
As compensation, more resources are required for SLO guarantees.
This contradicts the goal of Graft, i.e., resource-efficient serving on server sides.
Consequently, we focus on one partition point-based hybrid DL.
Additionally, finer-grained partitioning for DNNs with complex structures, such as parallel branches, has also been explored~\cite{infocom19-dads,sc22-ice}.
Yet, considering the trade-off between time complexity and resource efficiency, we treat parallel branches as a unit without further partitioning.


\ifCLASSOPTIONcaptionsoff
  \newpage
\fi

\bibliographystyle{IEEEtran}
\bibliography{refs-revise}

\begin{IEEEbiography}[{\includegraphics[width=1in,height=1.25in,clip,keepaspectratio]{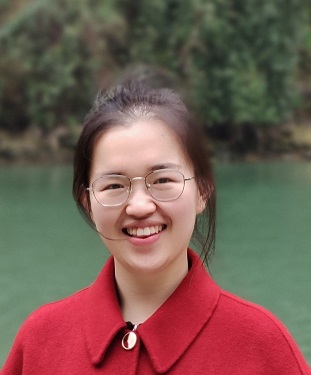}}]{Jing Wu} received the M.S. degree in the School of Computer Science and Engineering, Northeastern University, Shenyang, China, in 2018. She is currently a Ph.D. student in the
School of Computer Science and Technology,
Huazhong University of Science and Technology, China. Her research interests include
edge computing, augmented reality, and deep learning.
\end{IEEEbiography}

\begin{IEEEbiography}[{\includegraphics[width=1in,height=1.25in,clip,keepaspectratio]{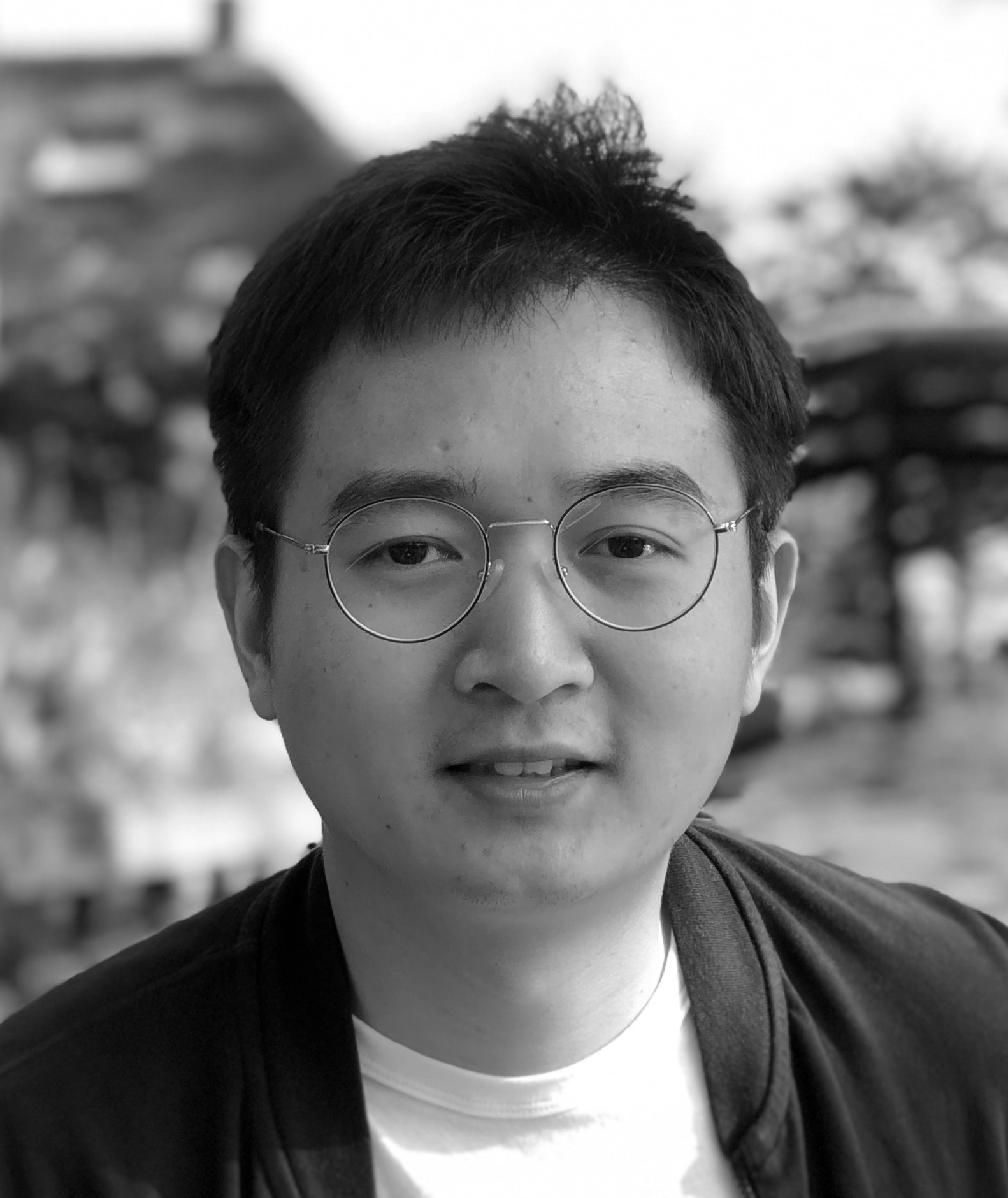}}]{Lin Wang} is currently a Full Professor and Head of the Chair of Computer Networks in the Department of Computer Science at Paderborn Univesity. He received his Ph.D. degree from the Institute of Computing Technology, Chinese Academy of Sciences in 2015. Previously, he held positions at Vrije Universiteit Amsterdam, TU Darmstadt, SnT Luxembourg, and IMDEA Networks Institute. He is broadly interested in networked systems, with a focus on in-network computing and intermittently-powered IoT systems. He has received a Google Research Scholar Award, an Outstanding Paper Award of RTSS 2022, a Best Paper Award of IPCCC 2023, and an Athene Young Investigator Award of TU Darmstadt. He is currently a Senior Member of IEEE. 
\end{IEEEbiography}

\begin{IEEEbiography}[{\includegraphics[width=1in,height=1.25in,clip,keepaspectratio]{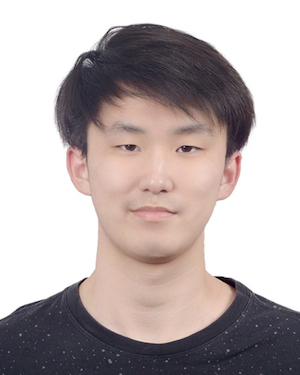}}]{Qirui Jin} is a current undergraduate student at the University of Michigan — Ann Arbor. His research topics include machine learning systems, reverse engineering, and neural network verification.
\end{IEEEbiography}

\begin{IEEEbiography}[{\includegraphics[width=1in,height=1.25in,clip,keepaspectratio]{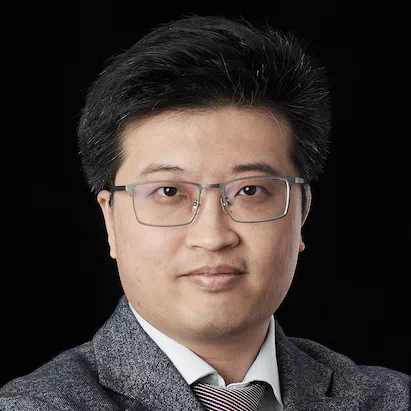}}]{Fangming Liu}
(S'08, M'11, SM'16) received the B.Eng. degree from the Tsinghua University, Beijing, and the Ph.D.	degree from the Hong Kong University of Science and Technology, Hong Kong. He is currently a Full Professor with the Huazhong University of Science and Technology, Wuhan, China. His research interests include cloud computing and edge computing, datacenter and green computing, SDN/NFV/5G and applied ML/AI. He received the National Natural Science Fund (NSFC) for Excellent Young Scholars, and the National Program Special Support for Top-Notch Young Professionals. He is a recipient of the Best Paper Award of IEEE/ACM IWQoS 2019, ACM e-Energy 2018 and IEEE GLOBECOM 2011, the First Class Prize of Natural Science of Ministry of Education in China, as well as the Second Class Prize of National Natural Science Award in China.
\end{IEEEbiography}

\end{document}